\documentclass[a4paper,fleqn,usenatbib,useAMS]{mnras}

\usepackage[T1]{fontenc}
\usepackage{ae,aecompl}

\usepackage{graphicx}	
\usepackage{amsmath}	
\usepackage{amssymb}	
\usepackage{float}
\usepackage{booktabs}
\usepackage{rotating}
\usepackage{natbib}
\usepackage{fmtcount}
\usepackage{threeparttablex}
\usepackage{multirow}
\usepackage{morefloats}
\usepackage{dblfloatfix}
\usepackage{dcolumn}
\usepackage{amsmath}
\usepackage{latexsym}
\usepackage{longtable}
\usepackage{lipsum} 
\usepackage{textcomp}
\usepackage{gensymb}
\usepackage{mathtools}
\usepackage{afterpage}

\title[SN 2016B: A transitional type II SN]{SN 2016B a.k.a ASASSN-16ab: a transitional type II supernova}
\author[Raya Dastidar et al.]{Raya Dastidar$^{1,2}$\thanks{E-mail: rayadastidar@aries.res.in, rdastidr@gmail.com},
Kuntal Misra$^{1,3}$, Mridweeka Singh$^{1,4}$,
D. K. Sahu$^{5}$,
\newauthor
 A. Pastorello$^{6}$,
Anjasha Gangopadhyay$^{1,4}$,
L. Tomasella$^{6}$,
S. Benetti$^{6}$,
\newauthor
G. Terreran$^{7}$,
Pankaj Sanwal$^{1,4}$,
Brijesh Kumar$^{1}$,
Avinash Singh$^{5,8}$,
\newauthor
Brajesh Kumar$^{5}$,
G. C. Anupama$^{5}$,
S. B. Pandey$^{1}$
\\
$^{1}$Aryabhatta Research Institute of observational sciencES, Manora Peak, Nainital 263 001 India\\
$^{2}$Department of Physics \& Astrophysics, University of Delhi, Delhi 110 007\\
$^{3}$Department of Physics, University of California, 1 Shields Ave, Davis, CA 95616-5270, USA\\
$^{4}$Pt.Ravi Shankar Shukla University, Raipur 492 010,  India\\
$^{5}$Indian Institute of Astrophysics, Koramangala, Bengaluru 560 034, India\\
$^{6}$INAF Osservatorio Astronomico di Padova, Vicolo dell'Osservatorio 5, 35122 Padova, Italy\\
$^{7}$Center for Interdisciplinary Exploration and Research in Astrophysics (CIERA) and Department of Physics and Astronomy, 
\\
Northwestern University, Evanston, IL 60208\\
$^{8}$Joint Astronomy Programme, Department of Physics, Indian Institute of Science, Bengaluru  560 012, India
}

\date{Accepted XXX. Received YYY; in original form ZZZ}

\pubyear{2018}

\begin{document}
\label{firstpage}
\pagerange{\pageref{firstpage}--\pageref{lastpage}}
\maketitle

\begin{abstract}
We present photometry, polarimetry and spectroscopy of the Type II supernova ASASSN-16ab/SN~2016B in PGC~037392. The photometric and spectroscopic follow-up commenced about two weeks after shock breakout and continued until nearly six months. The light curve of SN~2016B exhibits intermediate properties between those of Type IIP and IIL. The early decline is steep (1.68~$\pm$~0.10~mag~100~d$^{-1}$), followed by a shallower plateau phase (0.47~$\pm$~0.24 mag 100~d$^{-1}$). The optically thick phase lasts for 118~d, similar to Type IIP. The $^{56}$Ni mass estimated from the radioactive tail of the bolometric light curve is 0.082~$\pm$~0.019~M$_\odot$. High velocity component contributing to the absorption trough of H$\alpha$ and H$\beta$ in the photospheric spectra are identified from the spectral modelling from about 57 - 97~d after the outburst, suggesting a possible SN ejecta and circumstellar material interaction. Such high velocity features are common in the spectra of Type IIL supernovae. By modelling the true bolometric light curve of SN~2016B, we estimated a total ejected mass of $\sim$15~M$_\odot$, kinetic energy of $\sim$1.4~foe and an initial radius of $\sim$400~R$_\odot$.
\end{abstract}




\section{Introduction}
The classification of supernovae (SNe), from the 1940s to present, has been through numerous challenges. With the detailed study of events, their individual properties surfaced, which led to the proposal of various classes of SNe. Type II SNe are the result of the core-collapse of massive stars ($\gtrsim$ 8~M$_\odot$), showing abundant hydrogen in their spectra at peak. In order of increasing hydrogen envelope stripping, Type II SNe are classified into II-plateau (IIP), II-linear (IIL) and IIb. The hydrogen rich Type II SNe viz. IIP and IIL, are categorised based on the presence of a roughly constant luminosity phase or a linearly declining photospheric phase, respectively, in their optical light curve, at epochs post peak brightness \citep{1979A&A....72..287B}. 

There has been a long standing debate over classifying SNe II into Type IIP and IIL, which has recently escalated by sample studies of Type II SNe, some advocating while others disfavouring the classification. The studies of \cite{2012ApJ...756L..30A} and \cite{2014MNRAS.445..554F} prescribes to retain the historic classification, whereas \cite{2014ApJ...786...67A} and \cite{2015ApJ...799..208S} believe that the distinction stem from the paucity in observed Type~IIL SNe as well as the limited time for which these events are generally observed, due to the rapid decline in their brightness compared to SNe IIP. This led \cite{2014ApJ...786...67A} to propose a more reasonable nomenclature viz. classifying hydrogen-rich explosions simply as SNe~II with a mention of its plateau decline rate parameter. Observationally, strong P-Cygni profile of H {\sc i} lines in the spectra are the typical signatures of Type II SNe, while their light curves exhibit a range of morphologies. These differences possibly arise due to the varying mass and density profile of the hydrogen envelope of the progenitor star at the time of explosion or may be an effect of dense circumstellar material surrounding the progenitor \citep{2017ApJ...838...28M}.

While many theoretical studies are trying to provide a plausible explanation of the explosion mechanism of these events, observationally, the explosion geometry of SNe is a key diagnostic to constrain their explosion mechanism. However, owing to the unresolvability of extragalactic sources, the precise determination of ejecta structure and geometry becomes arduous. Nevertheless, the recent simulations of exploding stars, which require a multidimensional explosion scenario for a successful outburst \citep[e.g.][]{2015ApJ...801L..24M,2016PASA...33...48M,2016ApJ...831...98R}, is a definitive cue of the asymmetric nature of these explosions. The polarization of the emitted radiation from SNe may be attributed to this \citep{1996ApJ...459..307H,1999ApJ...521..179H,2003ApJ...592..457W,2006Natur.440..505L},
others being electron scattering in the SN ejecta \citep{1991ApJ...375..264J,2011MNRAS.415.3497D} and the reflection of light from the circumstellar and the interstellar material \citep{1986ApJ...308..225C,1996ApJ...467..435W,2017ApJ...834..118M}. Broadband polarimetry and spectropolarimetry prove to be handy to decode the imprints of polarization engraved in the emitted radiation \citep[see][for a review]{2008ARA&A..46..433W}. While the overall structure of the ejecta can be understood from the spectra of SNe, polarimetric studies reveal the shape, composition, opacity and ionisation state of the SN ejecta. 

Type IIP SNe are expected to show low levels of continuum polarization at early times ($\sim$0.1\%) and an increasing trend in polarization levels ($\sim$1\%) as the inner layers of the ejecta are revealed \citep{2001PASP..113..920L,2002PASP..114.1333L,2010ApJ...713.1363C,2016MNRAS.456.3157K,2017ApJ...834..118M}, which might stem from an aspherical core \citep{2005AstL...31..792C,2006AstL...32..739C}. However, \cite{2011MNRAS.415.3497D} and \cite{2016IAUFM..29B.458L} attributed the rise in polarization levels to the decreasing electron scattering opacity at late times. So the initial low polarization might be electron scattering optical depth effect, rather than arising from a spherical envelope. The early time asphericity, thus, gets subdued by the photons scattered off multiple times by the electrons, which in the process loses all its memory of the initial polarization state.  

In this paper, we present the results of the monitoring campaign of a Type II event, ASASSN-16ab/SN~2016B, in the galaxy PGC 037392. The paper is organized as follows: Section~\ref{sec2} presents the parameters of SN 2016B and in Section \ref{sec3} we discuss the data acquired and the reduction procedure. The estimation of total line of sight extinction is discussed in Section \ref{sec4}. The photometric evolution and light curve parameter estimates are discussed in Section \ref{sec5}.  The polarization measurements and the derived parameters are described in Section \ref{sec6}. The bolometric light curve, $^{56}$Ni mass and explosion parameters from analytical modelling of the bolometric light curve are derived in Section \ref{sec7}. We present the spectral evolution, {\sc syn++} modelling and derive line velocities and an estimate of the progenitor mass of SN 2016B in Section \ref{sec8}. In Section \ref{sec9}, we discuss the transitional nature of SN~2016B and summarize the results in Section \ref{sec10}.  

\section{SN 2016B}
\label{sec2}
SN 2016B, a.k.a ASASSN-16ab, was spotted in PGC 037392 by the All Sky Automated Survey for SuperNovae (ASAS-SN) program on 2016 January 3.62 UT at a visual magnitude of 14.7~mag \citep{2016ATel.8502....1S}. The SN was located at approximately 4$^{\prime\prime}$.6~S and 9$^{\prime\prime}$.8~W from the centre of the host galaxy. Considering the Tully-Fisher distance to the galaxy, which is 26.8 $\pm$ 4.9~Mpc \citep[][reported in NED]{2015AstBu..70....1K}, the deprojected radial distance of the SN from the host galaxy nucleus is 2.3 kpc. The SN was also detected by ASAS-SN in images obtained on 2015 December 27.5~UT at $V$$\sim$ 16.3~mag and the last non-detection to a magnitude limit of $V$\textgreater 16.7~mag was reported on 2015 December 24.5~UT. The intermediate epoch between the last non-detection and first detection tightly constrain the explosion epoch to 2015 December 26~UT (JD 2457382.5 $\pm$ 1.5). A classification spectrum was obtained on 2016 January 04.11 UT using the SPRAT spectrograph on the Liverpool Telescope located at Roque de los Muchachos \citep{2016ATel.8505....1P}, which is consistent with a pre-maximum Type II SN spectrum. The details of SN 2016B and its host galaxy are given in Table~\ref{tab:sn16B_PGC037392_detail}. 

\begin{table}
\centering
 \begin{minipage}{84mm}
\caption{Basic information on SN 2016B and the host galaxy PGC 037392. The host galaxy parameters are taken from NED.}
\begin{tabular}{@{}cc@{}}
\hline
Host galaxy & PGC 037392   \\
Galaxy type & Scd \\  
Redshift & 0.004293 $\pm$ 0.000077$^1$ \\ 
Major Diameter & 0.59 arcmin \\
Minor Diameter & 0.39 arcmin \\
Helio. Radial Velocity &  1287$\pm$23 km s$^{-1}$ \\
\hline
Offset from nucleus & 9$^{\prime\prime}$.8 W,4$^{\prime\prime}$.6 S \\
Distance & 26.8 $\pm$ 4.9 Mpc$^2$ \\
Total Extinction E(B-V) & 0.075 $\pm$ 0.017 mag$^3$ \\
SN type & II\\
Date of Discovery & 2457389.12 (JD) \\
Estimated date of explosion & 2457382.0 $\pm$ 1.2 (JD)$^3$\\
\hline 
\end{tabular}
\newline
$^1$ \citet{1995AJ....109..874B}
$^2$ Tully-Fisher method.        
$^3$ This paper.       
\label{tab:sn16B_PGC037392_detail}   
   \end{minipage}
\end{table}

We used the {\sc snid} code \citep{2007ApJ...666.1024B} to cross-correlate the spectrum obtained by us on 2016 January 7~UT to the library of spectral templates in {\sc snid} and constrain the explosion epoch. We perform the fit in the wavelength range 3500-6000~\AA{} since during early phase most of the spectral lines in SNe II lie in this range. We report the best three matches in Table~\ref{expl_epoch} along with the quality of fit \textquoteleft rlap\textquoteright{} parameter and obtain a mean value of 14.2 $\pm$ 2.0~d as the epoch of the spectrum from the explosion date. This yields 2015 December 24.7~UT (JD~2457381.2 $\pm$ 2.0) as the explosion epoch of SN~2016B. This is in good agreement to that obtained from non-detection. The consistency of the two methods has also been indicated in the works of \cite{2014ApJ...786...67A} and \cite{2017ApJ...850...89G}. The mean of these two values yields 2015 December 25.4~UT (JD 2457382.0 $\pm$ 1.2), which we have adopted as the explosion date of SN~2016B.

\begin{table}
 \begin{minipage}{84mm}
  \caption{The best three matches to the spectrum obtained on 2016 January 7 along with the rlap parameter to estimate the age of SN 2016B.}
  
  \begin{tabular}{@{}lccccc@{}}
  \hline
SN & rlap$^\dagger$ & Age since V$_{max}$ & Age since explosion  \\
      &                                 &       (days)                        &  (days)  \\
 \hline
SN 2004et  & 10.2 & -1.9 &  14.2 $\pm$3 \\
SN 2004et & 8.8 & -0.9 & 15.2 $\pm$3 \\
SN 1999em & 8.6 & -2.9 & 11.3 $\pm$ 5\\
\hline
Mean & & & 14.2 $\pm$ 2.0 \\
\hline				  
\end{tabular}			  
$^\dagger$Quality of fit
     \label{expl_epoch}
\end{minipage}
\end{table}

\section{Data acquisition and Reduction}
\label{sec3}
\subsection{Photometry and Spectroscopy}
Optical monitoring of SN~2016B was conducted using the telescopes listed in Table \ref{tab:details_instrument_detectors}, equipped with broadband {\it UBVRI} and {\it ugriz} filters from $\sim$ 13 to 188 d post explosion. A single epoch of observation was obtained on 465~d since the estimated explosion time in the $R$-band with the 3.6m Devasthal Optical Telescope (DOT). The {\it ugriz} magnitudes were converted to {\it UBVRI} magnitudes using the transformation equations in \cite{2006A&A...460..339J}. The SN was also observed from $\sim$9 to 23~d post explosion with the Ultra-Violet/Optical Telescope \citep[UVOT;][]{2005SSRv..120...95R} in three UV ({\it uvw2}: $\lambda_c$ = 1928\AA, {\it uvm2}: $\lambda_c$ = 2246\AA, {\it uvw1}: $\lambda_c$ = 2600\AA) and three optical ({\it u}, {\it b}, {\it v}) filters on The Neil Gehrels {\it Swift} Observatory \citep{2004ApJ...611.1005G}. The UV data reduction is discussed in Appendix \ref{phot} and the magnitudes are provided in Table \ref{uv_photometry}. The optical images were reduced after the initial cleaning steps as discussed in Appendix \ref{phot} and the differential magnitudes of the SN (presented in Table \ref{photometry}) were obtained using the nightly zero-points computed from the local standards generated in the SN field (see Fig. \ref{fig:local_std} and Table \ref{tab:local}). The spectroscopic observations were carried out at 18 epochs from the facilities listed in Table \ref{tab:details_instrument_detectors} and the log of spectroscopic observation is tabulated in Table \ref{tab:spectra_log}. The spectroscopic reduction procedure is detailed in Appendix \ref{spectro}.

\subsection{Polarimetry}
The broad-band polarimetric observations of SN 2016B in the $R$ band ($\lambda_{eff}$ = 6700 \AA) were made using the  ARIES Imaging Polarimeter \citep[AIMPOL,][]{2004BASI...32..159R} mounted at the 1.04m Sampurnanand Telescope (ST) at Manora Peak, Nainital. The observations were carried out at seven epochs during January to March 2016 using the liquid nitrogen cooled Tektronix 1024 $\times$ 1024 pixel$^2$ CCD camera. However, only the central 370~~$\times$~370~pixel$^2$ CCD region was used for the observation.  The pixel scale of the CCD in this setup is 1.48~arcsec/pixel which corresponds to an unvignetted field of view covering $\sim$8~arcmin diameter of sky. The stellar FWHM is found to be well-constrained within two to three pixels. 

A half-wave plate modulator and a Wollaston prism analyser are used inside AIMPOL, which produce the ordinary and extraordinary images separated by 28 pixels along the north-south direction on the sky plane. Multiple frames are acquired at each of the four half-wave plate positions (i.e. 0$\degree$, 22.5$\degree$, 45$\degree$ and 67.5$\degree$), which are then aligned and combined to increase the signal-to-noise ratio.  
The ratio of the difference between the intensities of the ordinary (I$_o$) and extraordinary (I$_e$) beams to the total intensity is given by
\begin{equation}
R(\alpha) = \frac{I_e/I_o-1}{I_e/I_o+1} = P cos(2\theta - 4\alpha)
\end{equation}
where $P$ is the fraction of the total light in the linearly polarized condition, $\theta$ is the position angle of the plane of polarization and $\alpha$ is the half-wave plate position angle. This ratio is denoted by normalised Stokes' parameters q(=Q/I), u(=U/I), q$_1$(=Q$_1$/I)  and u$_1$(=U$_1$/I) when the half-wave plate's fast axis is rotated by $\alpha$~=~0$\degree$, 22.5$\degree$, 45$\degree$ and 67.5$\degree$ respectively with respect to the reference axis. Ideally, $P$ and $\theta$ can be determined from the measurements of the first two plate positions. However, due to varying response of the system to the two orthogonal components and the dependence of the response of CCD on the position of its surface, a correction factor is multiplied with the ratio of the observed signal in the two images  (I$_e^\prime$/I$_{o}^\prime$) 
given by the following expression \citep{1998A&AS..128..369R}:
\begin{equation}
\frac{F_o}{F_e} = \Bigg[\frac{I'_o(0\degree)}{I'_e(45\degree)} \times \frac{I'_o(45\degree)}{I'_e(0\degree)} \times \frac{I'_o(22.5\degree)}{I'_e(67.5\degree)} \times \frac{I'_o(67.5\degree)}{I'_e(22.5\degree)}\Bigg]^{0.25}
\end{equation}
Applying this correction factor, we derive the ratio at the four plate positions and fitting the cosine curve, the values of $P$ and $\theta$ are obtained. The errors in the measured ratios are limited by photon statistics and is given by \citep{1998A&AS..128..369R},
\begin{equation}
\sigma_{R(\alpha)} = \frac{\sqrt{N_e + N_o + N_{Be} + N_{Bo}}}{(N_e + N_o)},
\end{equation}
where $N_o$ and $N_e$ are ordinary and extraordinary counts and $N_{Bo}$ and $N_{Be}$ are ordinary and extraordinary background counts, respectively, for a position $\alpha$.
This error is then propagated to derive the errors in the polarization parameters. 

We observed two polarized standards \citep[HD 19820 and HD 25443, from][]{1992AJ....104.1563S} for calibration of the zero-point polarization angle, which mainly arises from the imperfect alignment of the fast axis of half-wave plate with the north-south direction. The measured degree of polarization~($P$) of the polarized standards are listed in Table \ref{tab:pol_std_stars} and were found to be consistent within error with the standard values and any deviation from the standard values of the polarization angle~($\theta$) is applied to the SN. The instrumental polarization of ST has been characterized since 2004 by observations of various sources \citep{2008MNRAS.388..105M, 2009MNRAS.396.1004P, 2013A&A...556A..65E,2014MNRAS.442....2K,2016MNRAS.457.1000S} and is found to be $\sim$0.1 per cent in $UBVRI$ bands. We applied this correction in our SN polarization measurements.

\begin{figure*}
		\includegraphics[scale=1.00, width=1.0\textwidth,clip, trim={4.8cm 1.6cm 4.6cm 3.0cm}]{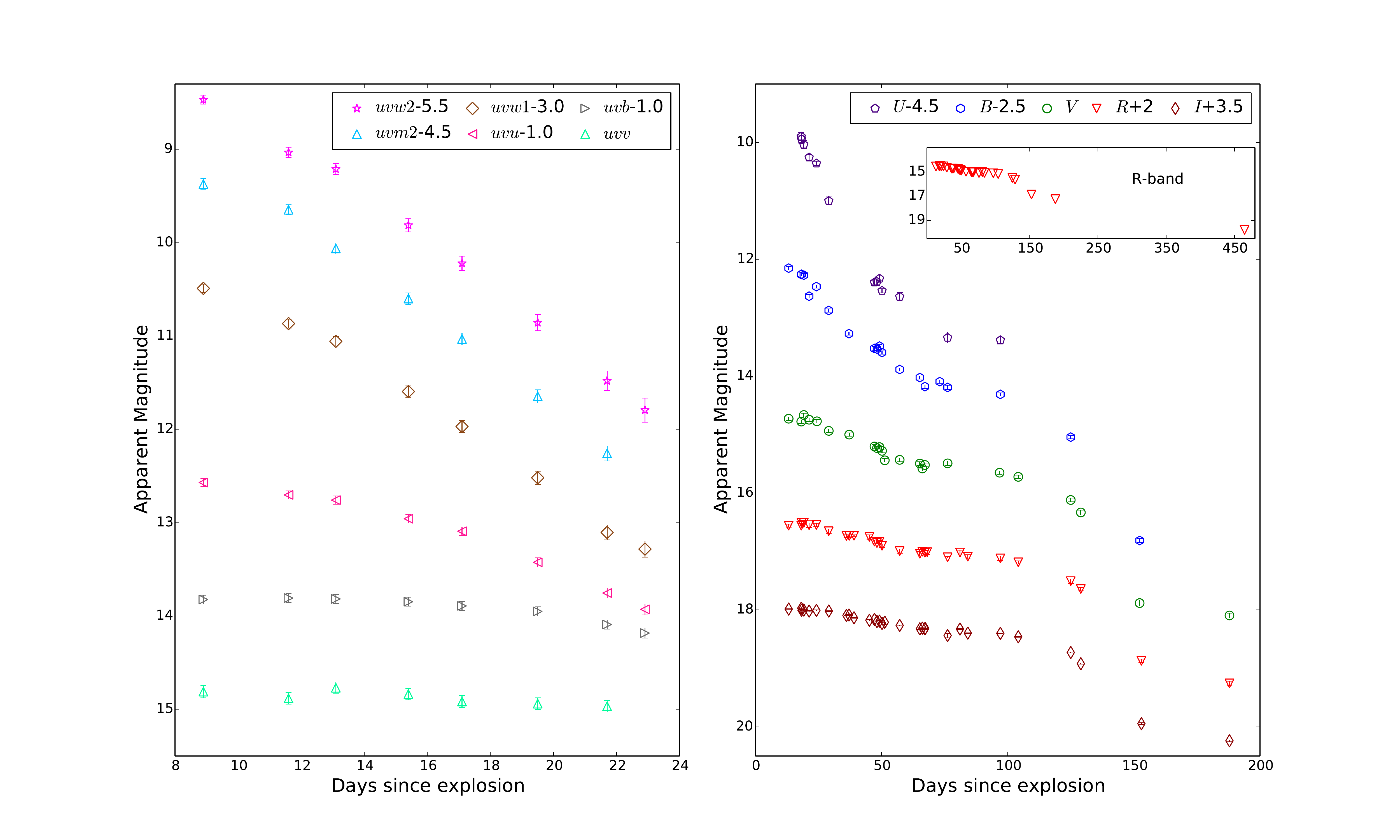}
	\caption{{\it Swift} UVOT (left panel) and broadband {\it UBVRI} (right panel) light curves of SN 2016B shifted arbitrarily for clarity. The inset plot in the right panel shows the $R$-band light curve up to the last point obtained on 465~d with the 3.6m DOT.}
	\label{fig:light_curve}
\end{figure*}

\section{Extinction}
\label{sec4}
The intrinsic nature of an event is unveiled once the reddening contributions from the Milky~Way and the host galaxy along its line of sight are discerned. The \cite{2011ApJ...737..103S} IR based dust map estimates a colour excess E(B$-$V) of 0.0180 $\pm$ 0.0012~mag for Galactic extinction in the direction of SN 2016B. 

To estimate host galaxy extinction, use of empirical relations correlating the equivalent width of Na {\sc i} D narrow absorption line to the colour excess has been done for a number of SNe, even though it has been shown to be a poor indicator \citep{2013ApJ...779...38P}. The rest frame low-resolution spectra of SN 2016B, however, does not exhibit any conspicuous narrow absorption feature around 5893 \AA, possibly indicating low levels of host galaxy extinction. This agrees with the remote location of the SN in its host galaxy, free of any nearby contaminating sources.

Consequently, we resorted to the \textquotedblleft colour method\textquotedblright{} \citep{2010ApJ...715..833O} to estimate the host galaxy extinction in SN~2016B. In this method, the opacity of SNe IIP is assumed to be dominated by electron scattering. Since all SNe IIP will attain nearly the same recombination temperature at the end of the plateau phase, these ought to have similar intrinsic colours at this phase \citep{2002ApJ...566L..63H}. Hence, the different colours exhibited by SNe~IIP at these epochs may be attributed to reddening due to dust along its line of sight.
The conversion factor computed by \cite{2010ApJ...715..833O} between E(V$-$I) and A${_V}$ using a library of SN II spectra is 2.518 and the following prescription is suggested:
\begin{gather}
A_V(V-I) = 2.518[(V-I)-0.656]\\
\sigma(A_V) = 2.518 \sqrt{\sigma_{(V-I)} + 0.0053^2 + 0.0059^2}
\end{gather}
We calculate the weighted mean of $(V-I)$ colours (corrected for Galactic extinction) on 97.3~d and 104.4~d from the explosion corresponding to the end of the plateau phase, which results in $(V-I)$ = 0.726 $\pm$ 0.021 mag. The estimated $A_{V(host)}$ is 0.176 $\pm$ 0.054~mag and with a total-to-selective extinction ratio (R$_V$) of 3.1, we obtain E(B$-$V)$_{host}$ = 0.057 $\pm$ 0.017~mag.  
Hence, the total reddening to the SN, considering the contributions from both the Milky Way and host galaxy reddening is E(B$-$V)$_{tot}$~=~0.075~$\pm$~0.017~mag, which we have adopted throughout the paper.

\section{Photometric evolution of SN 2016B}
\label{sec5}

The light curve slope and luminosities of the early declining part, plateau phase and radioactive tail phase provide clues to our understanding of the nature and energetics of the explosion. The decline rates during these phases indicate the extent of thermalization in the ejecta, with shallow slopes of the plateau and the radioactive tail suggesting massive ejecta retained by the progenitor star \citep{2004ApJ...617.1233Y}. To examine the behaviour of SN~2016B, we concocted a comparison sample of SNe with a variety of early decline rates in $V$-band; SNe~2004et \citep{2006MNRAS.372.1315S,2007MNRAS.381..280M,2010MNRAS.404..981M}, 2007od \citep{2011MNRAS.417..261I}, 2009bw \citep{2012MNRAS.422.1122I}, 2009dd \citep{2013A&A...555A.142I}, 2013ej \citep{2015ApJ...807...59H,2015ApJ...806..160B,2016MNRAS.461.2003Y,2016ApJ...822....6D}, and 2016esw \citep{2018MNRAS.478.3776D} have decline rates between 0.5 to 1.0~mag in the first 50~d after maximum; SN~2013by \citep{2015MNRAS.448.2608V}, a typical SN IIL, exhibits the highest decline rate (1.5~mag in 50~d) while SN 1999em \citep{2001ApJ...558..615H,2002PASP..114...35L,2003ApJ...594..247L} displays the lowest decline rate~(0.25 mag in 50~d). We also include the peculiar Type II SN~1987A \citep{1990AJ.....99.1146H} in the comparison sample.

\begin{figure}
		\includegraphics[scale=1.00, width=0.52\textwidth,clip, trim={0.8cm 9.7cm 0cm 1.8cm}]{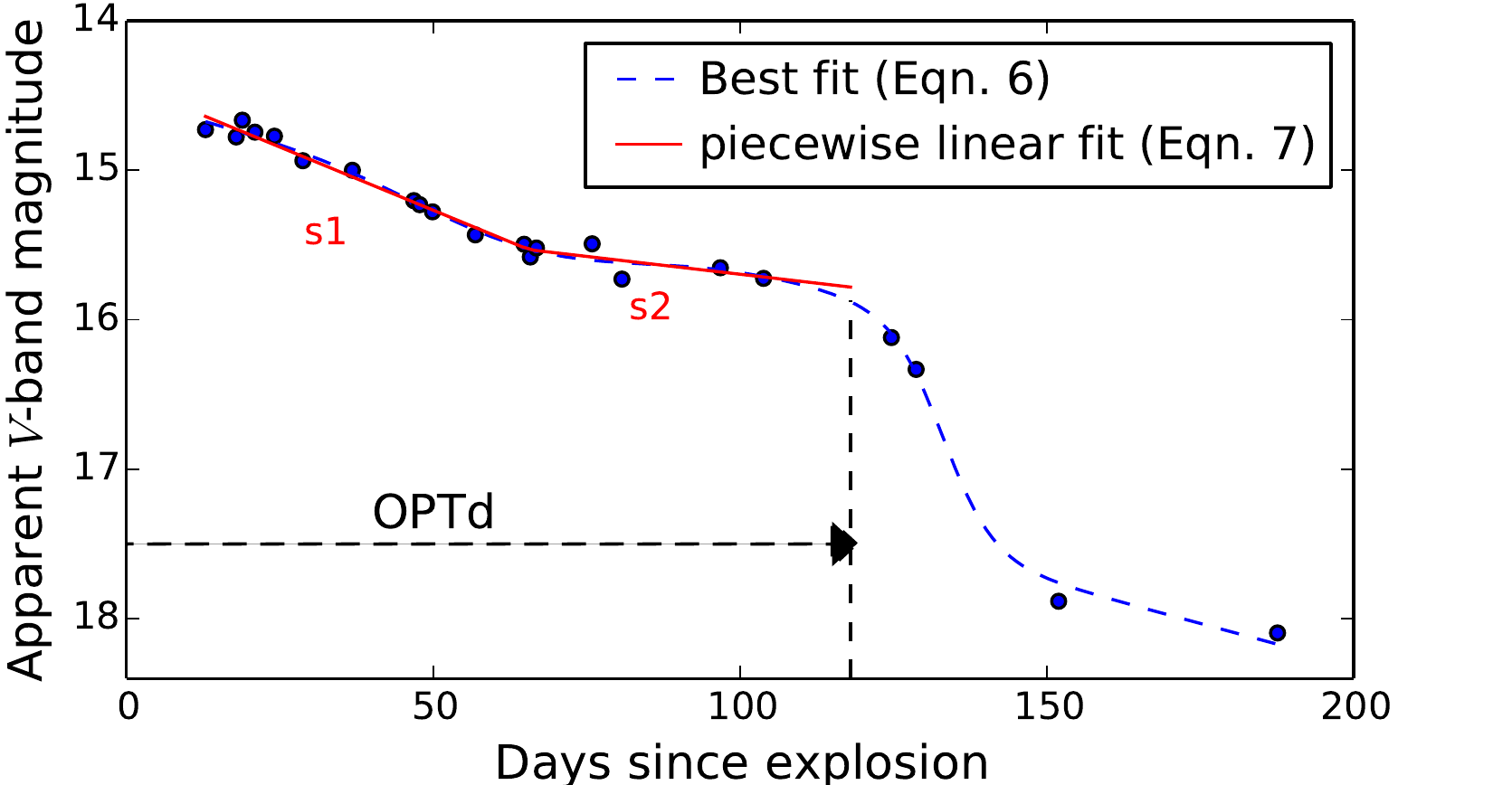}
	\caption{The $V$-band light curve of SN 2016B fitted with the expression given in \citet{2010ApJ...715..833O} shown with the (blue) dashed lines. The red lines are from the fit with the piecewise linear function to the light curve with an early declining phase (s$_1$) and a plateau phase (s$_2$). The dashed vertical line denotes the end of the plateau phase.}
	\label{fig:fitted_light_curve}
\end{figure}

The UV and optical light curves of SN~2016B are presented in Fig. \ref{fig:light_curve}. The observed multi-band light curves of SN~2016B commences with a decline, which characterizes the initial expansion and cooling of the ejecta. The UV light curve steeply decline in the first 25~d at a rate of 0.20~$\pm$~0.01~mag~d$^{-1}$ in $uvw1$ and 0.23~$\pm$~0.01~mag~d$^{-1}$ in both $uvw2$ and $uvm2$ bands. The $UBVRI$ light curve then enters the plateau phase supported by the heating from the recombination of the initially shock ionized hydrogen. At the end of the plateau phase, the SN luminosity starts to decay in the tail phase which is powered by the instantaneous energy deposition from the decay of $^{56}$Co to $^{56}$Fe. 

In order to determine the light curve parameters, we fit the $V$-band light curve with the equation given in \cite{2010ApJ...715..833O}:
\begin{equation}
\label{equ6}
y(t) = \frac{-a_0}{1+e^{(t-t_{PT})/w_0}} + p_0 \times (t-t_{PT}) + m_0 - P e^{-(\frac{t-Q}{R})^2}
\end{equation}
where {\it t} is the time since explosion in days, {\it t$_{PT}$} is the time in days from explosion to the transition point between the end of the plateau and start of the radioactive tail phase, {\it a$_0$}~is the depth of the drop from plateau to nebular phase, {\it w$_0$} assesses the width of the transition phase in days, {\it p$_0$}~constrains the slope of radioactive tail, $P$ is the height of the Gaussian peak in units of magnitude, $Q$ is the centre of the Gaussian function in days, and $R$ is the width of the Gaussian function. From the best fit (as shown in Fig. \ref{fig:fitted_light_curve}), we obtain {\it t$_{PT}$}~=~(133.5~$\pm$~2.6)~d and the magnitude drop from plateau to tail phase ($a_0$) to be 1.6 $\pm$ 0.2~mag in a span of 25~d.

\begin{table*}
\caption{Parameters of the SNe II sample.}
\scalebox{0.8}{

\begin{tabular}{@{}llllllllllll}
\hline\hline
\\
&Parent & Distance$^\dagger$ & $A_V^{tot}$ & M$^{V}$ & OPTd & E (10$^{51})$ & R & M$_{ej}$  & $^{56}$Ni & Ref.\\

Supernova& Galaxy & (Mpc) & (mag) & (mag) & (days) & (ergs) & (R$_{\odot}$) & (M$_{\odot}$) &(M$_{\odot}$)\\

\hline
&&&&&&&&&&\\
1987A & LMC & 0.05 & 0.60 & - & 40 & 1.3 & 40 & 15 & 0.075 & 1\\
&&&&&&&&&&\\
1999em & NGC 1637 & 11.7 (0.1) & 0.31 & $-$16.71 & 95 & 1.2$^{+0.6}_{-0.3}$ & 249$^{+243}_{-150}$ & 27$^{+14}_{-8}$ & 0.042$^{+0.027}_{-0.019}$ & 2,3,4\\
&&&&&&&&&&\\
2004et &  NGC 6946 &  5.4 (1.0) &  1.27 &  $-$17.04 &  110$\pm$10 &  0.98$\pm$0.25 & 530$\pm$280 &  16$\pm$5 & 0.06$\pm$0.03 & 5,6,7\\
&&&&&&&&&&\\
2007od & UGC~12846 & 25.3 (0.8) & 0.12 & $-$17.64 & 25 & 0.5 & 670 & 5-7 & 0.02 & 8\\
&&&&&&&&&&\\
2009bw & UGC~2890 & 20.2 (0.6) & 0.96 & $-$17.24 & 100 & 0.3 & 510-1000 & 8.3-12 & 0.022 & 9\\
&&&&&&&&&&\\
2009dd & NGC~4088 & 14.0 (0.4) & 1.40 & $-$16.6 & $\sim$100 & 0.2 & 719 & 8.0 & 0.029 & 10\\
&&&&&&&&&&\\
2013by & ESO~138-G10 & 14.74 (1.0) & 0.60 & $-$17.12 (0.11) & 75 & - & - & - & 0.029 & 11\\
&&&&&&&&&&\\
2013ej & NGC~628 & 9.6 (0.7) & 0.19 & $-$16.61(0.10) & 85 & 2.3 & 450 & 12 & 0.020 & 12\\
&&&&&&&&&&\\
2016esw & UGC~2890 & 118.1 & 0.74 & $-$17.24 & 105 & 1.0 & 190 & -  & - & 13\\
&&&&&&&&&&\\
{\bf 2016B} & {\bf PGC~037392} & {\bf 26.8 (4.9)} & {\bf 0.23} & {\bf $-$16.80} & {\bf 118} & {\bf 1.4} & {\bf 400} & {\bf 14.7} & {\bf 0.082 $\pm$ 0.019} & {\bf This paper}\\
&&&&&&&&&&\\
\hline
\end{tabular}}
\label{parameter_SNII_sample}
\flushleft
$^\dagger$ In the H$_0$ = 73.24 km s$^{-1}$ Mpc$^{-1}$ scale \citep{2016ApJ...826...56R}.\\
References: (1) \cite{1990AJ.....99.1146H},
(2) \cite{2001ApJ...558..615H},
(3) \cite{2002PASP..114...35L},
(4) \cite{2003ApJ...594..247L},
(5) \cite{2006MNRAS.372.1315S},
(6) \cite{2007MNRAS.381..280M},
(7) \cite{2010MNRAS.404..981M},
(8) \cite{2011MNRAS.417..261I},
(9) \cite{2012MNRAS.422.1122I},
(10) \cite{2013A&A...555A.142I},
(11) \cite{2015MNRAS.448.2608V},
(12) \cite{2015ApJ...807...59H,2015ApJ...806..160B,2016MNRAS.461.2003Y,2016ApJ...822....6D},
(13) \cite{2018MNRAS.478.3776D}.
\end{table*}

\begin{figure}
	\begin{center}
		\includegraphics[scale=1.00, width=0.5\textwidth,clip, trim={0.2cm 0.2cm 0cm 1.3cm}]{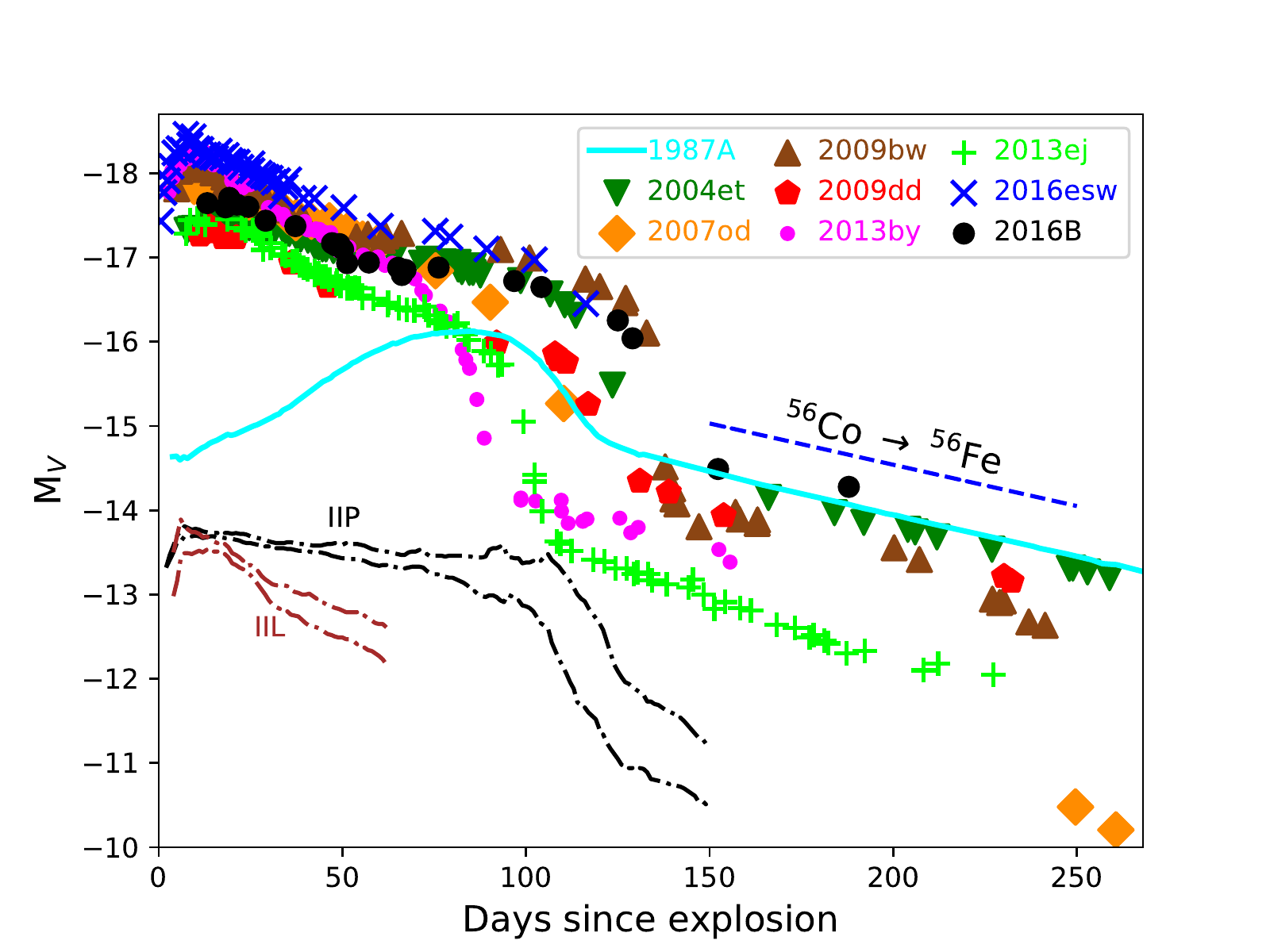}
	\end{center}
	\caption{Comparison of absolute $V$-band light curves of SN 2016B with other Type II SNe. The magnitudes are corrected for distance and reddening as listed in Table \ref{parameter_SNII_sample}. The radioactive decay line assuming full trapping of photons is shown with dashed lines. The dash-dotted lines depicts the range of slopes of Type IIP and IIL SNe as presented in \citet{2014MNRAS.445..554F}.}
	\label{fig:abs_light_curve}
\end{figure}

We apply the distance and reddening corrections to derive the $V$-band absolute magnitudes of SN 2016B and other SNe II (see Fig. \ref{fig:abs_light_curve}). The distance and extinction values for the comparison sample are listed in Table \ref{parameter_SNII_sample}. The absolute $V$-band light curve of SN 2016B shows a gradual decline up to about 66~d ($\sim$0.84 $\pm$ 0.05~mag~50~d$^{-1}$), thereafter it settles on to a roughly constant magnitude phase lasting up to 118~d since explosion (0.24 $\pm$ 0.12~mag~50~d$^{-1}$). These values are estimated by fitting a smoothed piecewise linear function to the apparent $V$-band magnitudes:
\begin{equation}
\label{equ7}
m_v = a + a_1t + a_2\Bigg(\frac{t-T_{tran}}{2} + \sqrt{\frac{(t-T_{tran})^2}{4} + b}\Bigg)
\end{equation}
The break time ($T_{tran}$) is the transition time from the early declining phase to the plateau phase, $a_1$ is the slope of the early declining phase, ($a_1$+$a_2$) is the slope of the plateau phase and $b$ is the smoothing factor. The two slopes ($s_1$ and $s_2$) from the best fit are shown in Fig. \ref{fig:fitted_light_curve} and could possibly be arising from a density discontinuity in the ejecta of SN~2016B. The optically thick phase duration (OPTd), defined in \cite{2014ApJ...786...67A} as the time span from explosion until the extrapolated plateau is more luminous than the light curve by 0.1~mag, is $\sim$118~d (marked with a vertical dashed line in Fig. \ref{fig:fitted_light_curve}). We fit both equation (\ref{equ6}) and (\ref{equ7}) on the $V$-band light curve of the comparison sample and list the parameters for the sample in Table~\ref{lc_par_comp}.

\begin{table*}
  \caption{Best fit values of t$_{PT}$, a$_0$ using \citep{2010ApJ...715..833O} expression and s$_1$, s$_2$ using smoothed piecewise linear function.}
  \label{lc_par_comp}
  \begin{tabular}{@{}lccccccc@{}}
  \hline
  \hline
SN &  t$_0$ & t$_0$-t$_{PT}$ & a$_0$ & s$_1$ & s$_2$ & T$_{tran}$ & References \\
      &  JD (2400000+)&  (days)     & (mag) &          &    \\
  \hline
SN 1999em &  51477.0 (1.0) & 112.6 $\pm$ 1.4 & 2.20 $\pm$ 0.17 & 0.50 $\pm$ 0.06 & -  & - & 1,2\\
SN 2004et & 53270.5 (0.2) & 123.9 $\pm$ 3.4 & 2.11 $\pm$ 0.06 & 1.27 $\pm$ 0.14 & 0.74 $\pm$ 0.15 & 43 $\pm$ 5 & 3,4,5\\
SN 2009bw & 54916.5 (3.0) & 135.3 $\pm$ 2.5 & 2.31 $\pm$ 0.23 & 1.90 $\pm$ 0.07 & 0.43 $\pm$ 0.17 & 49 $\pm$ 5 & 6\\
SN 2013by &  56404.0 (2.0) & 86.5 $\pm$ 0.4 & 2.56 $\pm$ 0.23 & 3.47 $\pm$ 0.20 & 2.35 $\pm$ 0.24 & 28 $\pm$ 2 & 7\\
SN 2013ej &  56497.5 (0.3) & 99.7 $\pm$ 0.3 & 2.82 $\pm$ 0.04 & 3.21 $\pm$ 0.28 & 1.63 $\pm$ 0.33 & 41 $\pm$ 2 & 8,9,10\\
SN 2016esw & 57608.3 (0.2) & - & - & 1.83 $\pm$ 0.06 & 1.02 $\pm$ 0.27 & 60 $\pm$ 9 &11\\
SN 2016B & 57382.0 (1.2) & 133.5 $\pm$ 2.6 & 1.6 $\pm$ 0.2 & 1.68 $\pm$ 0.10 & 0.47 $\pm$ 0.24 & 66 $\pm$ 5 & This paper\\

\hline 
\end{tabular}	

References: (1) \cite{2001ApJ...558..615H}, (2) \cite{2002PASP..114...35L},
(3) \cite{2006MNRAS.372.1315S}
(4) \cite{2007MNRAS.381..280M},
(5) \cite{2010MNRAS.404..981M},
(6) \cite{2012MNRAS.422.1122I},
(7) \cite{2015MNRAS.448.2608V},
(8) \cite{2015ApJ...806..160B},
(9) \cite{2016ApJ...822....6D},
(10) \cite{2016MNRAS.461.2003Y},
(11) \cite{2018MNRAS.478.3776D}.
\end{table*}

The effective plateau duration of SN 2016B is very short, however, we note that the OPTd is longer than any other SNe~IIL and is similar to SNe~IIP (usually lasting for more than 80~d in SNe IIP). The evolution is similar to SN 2009bw, whose early decline rate (0.95 mag 50~d$^{-1}$) is higher than SN~2016B by $\sim$0.1~mag, but has a similar plateau length and plateau slope. SN 2016esw show similar duration of the early declining phase lasting nearly 60~d, however, the slopes are steeper than SN 2016B. The light curve of SN 2016B becomes flatter thereafter, with a slope similar to the archetypal Type IIP SN~1999em (0.25~mag~50~d$^{-1}$), unlike SN~2013ej which continues to decline at a steady rate until it transitions to the radioactive tail.

While the light curve parameters, viz. the slope of the early cooling phase (s$_1$=1.68 $\pm$ 0.10~mag~100~d$^{-1}$) and the plateau slope (s$_2$=0.47 $\pm$ 0.24~mag~100~d$^{-1}$) of SN 2016B, are consistent with those of SNe II derived by \cite{2014ApJ...786...67A} (s$_1$ = 2.65 $\pm$ 1.50 and s$_2$ = 1.27 $\pm$ 0.93~mag~100~d$^{-1}$) within the errors, SN 2016B has a higher OPTd value. Further, we note that the decline rate of SN 2016B in $V$ and $R$-bands in the first 50~d (0.84~mag and 0.52~mag, respectively), requires SN 2016B to be classified as SN IIL \citep{2014MNRAS.445..554F,2011MNRAS.412.1441L}. Moreover, the early decline rates in $B$-band, which is 3.88~mag~100d$^{-1}$ \citep[\textgreater 3.5 mag~100~d$^{-1}$ for SNe IIL,][]{1994A&A...282..731P}, place SN 2016B in the IIL sub-class. Despite the overall similarity of the $V$-band light curve to the SN II family, the correlation found by \cite{2014ApJ...786...67A} between the absolute peak magnitude and the slope of the plateau (M$_V$ = $-$1.12$\times$$s_2$$-$15.99 mag) is not followed by SN 2016B. For the observed maximum absolute magnitude M$_V$ = $-$17.7~mag
(which may not be peak magnitude), the correlation predicts plateau slope greater than 1.5 mag~100~d$^{-1}$. This is significantly higher than the observed slope of 0.47~mag~100~d$^{-1}$. SN~2016esw was also found to deviate from the correlation, leading \cite{2018MNRAS.478.3776D} to suggest important implications for cosmology with SNe II.  

\begin{figure}
		\includegraphics[scale=1.00, width=0.53\textwidth,clip, trim={0.5cm 0.7cm 0cm 0cm}]{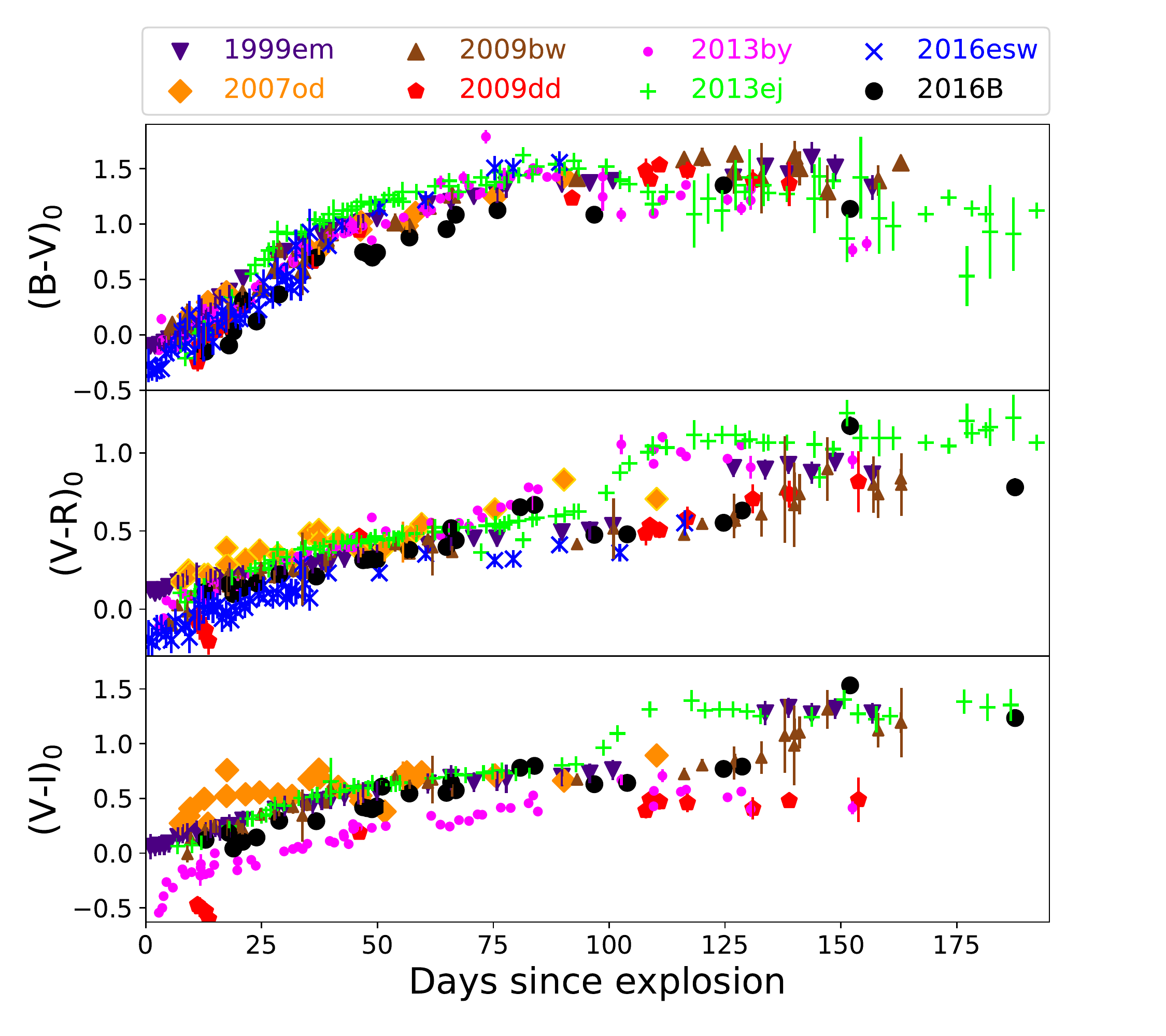}
	\caption{The (B-V)$_0$, (V-R)$_0$ and (V-I)$_0$ colour evolution of SN 2016B compared with other SNe II.}
	\label{fig:colour_light_curve}
\end{figure}

The temporal evolution of the reddening-corrected broadband colours serves as an important piece of information to understand the dynamics of SN ejecta. The reddening-corrected colour evolution [($B-V$)$_0$, ($V-R$)$_0$ and ($V-I$)$_0$] of SN 2016B is shown in Fig. \ref{fig:colour_light_curve}, with those of other SNe from the comparison sample. The ($B-V$)$_0$ colour becomes redder by 0.6 mag in 50 d, with indicates the cooling of the ejecta. In fact, this trend is seen in the ($V-R$)$_0$ and ($V-I$)$_0$ colours, which become redder at a much slower rate. This is because of the slowly evolving magnitude in the $VRI$ bands at early phase. While the ($B-V$)$_0$ colour of SNe~2013by and 2013ej, shows a bluer trend in the nebular phase, we do not find any perceptible trend in SN 2016B, similar to SNe 2009bw and 2009dd. Nevertheless, the overall evolution of SN 2016B is similar to most of comparison SNe II.

\section{Broad-band polarimetry}
\label{sec6}
Polarimetric observations of SN~2016B in the $R$-band have been carried out at 7 epochs during 16.4 - 73.5~d from explosion. The intrinsic polarization of the SN is contaminated by polarization of the SN light induced by aspherical dust grains along the SN line of sight. This additive component, the interstellar polarization (ISP), should be eliminated from the observed SN polarization value before claiming any polarization intrinsic to the SN. The ISP is formed by two components along the line of sight: polarization induced by the dust grains in the Milky Way (MW) and the host galaxy. However, there is no well-established method to determine these components as most of these methods rely on certain assumptions which are not always fulfilled. Below we discuss the methods used to estimate the MW and the host galaxy polarization. 
\subsection{Estimation of Milky Way polarization}
For SN 2016B, we determine the interstellar polarization using the field stars, observed in the same frame as SN 2016B. We selected three bright field stars and list their polarization parameters and distance \citep[taken from $Gaia$ DR2,][]{2016A&A...595A...1G,2018A&A...616A...1G} in Table \ref{tab:ISP_stars}. The ISP is computed from the observed stars by averaging over the Stokes parameters as follows: 
\begin{align}
\begin{split}
  \langle Q \rangle = \langle P_i cos(2\theta_i)\rangle_{i=1..n} \\
  \langle U \rangle = \langle P_i sin(2\theta_i)\rangle_{i=1..n},
\end{split}
\end{align}
which are then converted back to polarization parameters using:
\begin{align}
\begin{split}
P_{MW} = \sqrt{\langle Q \rangle^2 + \langle U \rangle^2}; \\
\theta_{MW} = \frac{1}{2}tan^{-1}\frac{\langle U \rangle}{\langle Q \rangle}
\end{split}
\end{align}
The mean ISP$_{MW}$ parameters, we obtain, are P$_{MW}$ = 0.43 $\pm$ 0.04~per~cent and $\theta_{MW}$ = (83 $\pm$ 4)\degree{} for SN~2016B.

\begin{table*}
 \centering
 \begin{minipage}{130mm}
\caption{Polarization measurements in the $R$-band for standard stars.}
\begin{tabular}{cccccccc}
\hline
Star ID & \multicolumn{2}{c}{Standard$^\dagger$} & UT Date & \multicolumn{2}{c}{Observed}\\
              & $P_R$(\%) & $\theta_R$($\degree$) & (yyyy-mm-dd)  & $P_R$(\%) & $\theta_R$($\degree$)\\
\hline 
\multirow{ 4}{*}{HD 19820} & \multirow{ 4}{*}{4.53 $\pm$ 0.03} & \multirow{ 4}{*}{114.46 $\pm$ 0.16} & 2016-01-10.9 & 4.176$\pm$0.166 & 123.167 $\pm$ 1.141\\ 
& & & 2016-01-15.0 & 4.414 $\pm$ 0.226 & 124.184 $\pm$ 1.465\\
& & & 2016-01-16.0 & 4.590 $\pm$ 0.138 & 126.442 $\pm$ 0.860\\
& & & 2016-02-08.0 & 4.396 $\pm$ 0.113 & 126.863 $\pm$ 0.736\\
\hline
\multirow{ 6}{*}{HD 25443} & \multirow{ 6}{*}{4.73 $\pm$ 0.32} & \multirow{ 6}{*}{133.65 $\pm$ 0.28} & 2016-01-10.9 & 4.408 $\pm$0.116 & 142.629 $\pm$ 0.756\\ 
& & & 2016-01-13.0 & 4.658 $\pm$ 0.142 & 143.230 $\pm$ 0.875\\
& & & 2016-01-15.0 & 5.003 $\pm$ 0.117 & 143.012 $\pm$ 0.673\\
& & & 2016-01-16.0 & 4.582 $\pm$ 0.133 & 143.265 $\pm$ 0.828\\
& & & 2016-02-09.0 & 4.696 $\pm$ 0.143 & 138.343 $\pm$ 0.872\\
& & & 2016-03-08.0 & 4.720 $\pm$ 0.074 & 143.255 $\pm$ 0.452\\
\hline
\end{tabular}
$^\dagger$ taken from \citet{1992AJ....104.1563S}
\label{tab:pol_std_stars}   
\end{minipage}
\end{table*}

\begin{table}
\setlength{\tabcolsep}{2pt}
\caption{Polarization measurements in the $R$-band of the field stars as marked in Fig. \ref{fig:local_std} towards the direction of SN 2016B.}
\begin{tabular}{cccccccc}
\hline
\#id & $\alpha_{J2000.0}$ & $\delta_{J2000.0}$ & P$_R$ & $\theta_R$ & Distance\\
 & (hh:mm:ss) &(dd:mm:ss) & per cent & ($\degree$)  &  pc\\
\hline 
E & 11:55:16.7 & +01:44:18.5 & 0.28 $\pm$ 0.08 & 75.4 $\pm$ 7.9 & 679 \\
G & 11:55:15.0 & +01:41:44.8 & 0.33 $\pm$ 0.14 & 80.8 $\pm$ 8.8 & 900 \\
I & 11:55:11.1 & +01:44:02.6 & 0.25 $\pm$ 0.06 & 90.5 $\pm$ 4.9 & 193 \\
\hline
\end{tabular}
\label{tab:ISP_stars}   
\end{table}

\subsection{Estimation of host galaxy polarization}
Estimating the host galaxy polarization is a non-trivial task. While now we know the Galactic dust properties, the characteristics of dust in the host galaxy remain largely unknown. These studies are possible in nearby galaxies, with two known cases of SNe 1986G \citep{1987MNRAS.227P...1H} and 2001el \citep{2003ApJ...591.1110W}, where it was inferred that the dust grain sizes are smaller than Galactic ones. Assuming that the dust grain properties of PGC 037392 is similar to Galactic dust, we roughly estimate the maximum degree of polarization due to host galaxy following \cite{1975ApJ...196..261S} relation, P$_{max}$ \textless 9 $\times$ E($B-V$). For extinction values of host galaxy estimated in Section \ref{sec3}, we obtain P$_{max}$ (for host) to be \textless0.68\% for SN 2016B.
Next, we require the magnetic field direction information for the host, to perform the subtraction of this component from the total component. It is well-established that the magnetic field runs parallel to the spiral arms in spiral galaxies. However, even being a Scd type galaxy, the spiral arms of PGC 037392 are not well-resolved, neither the position of the SN, given the large inclination of the galaxy. So, we restrain from subtracting host galaxy component from the total polarization.

\begin{figure}
		\includegraphics[scale=1.00, width=0.48\textwidth,clip, trim={0.4cm 0.3cm 1.3cm 1.3cm}]{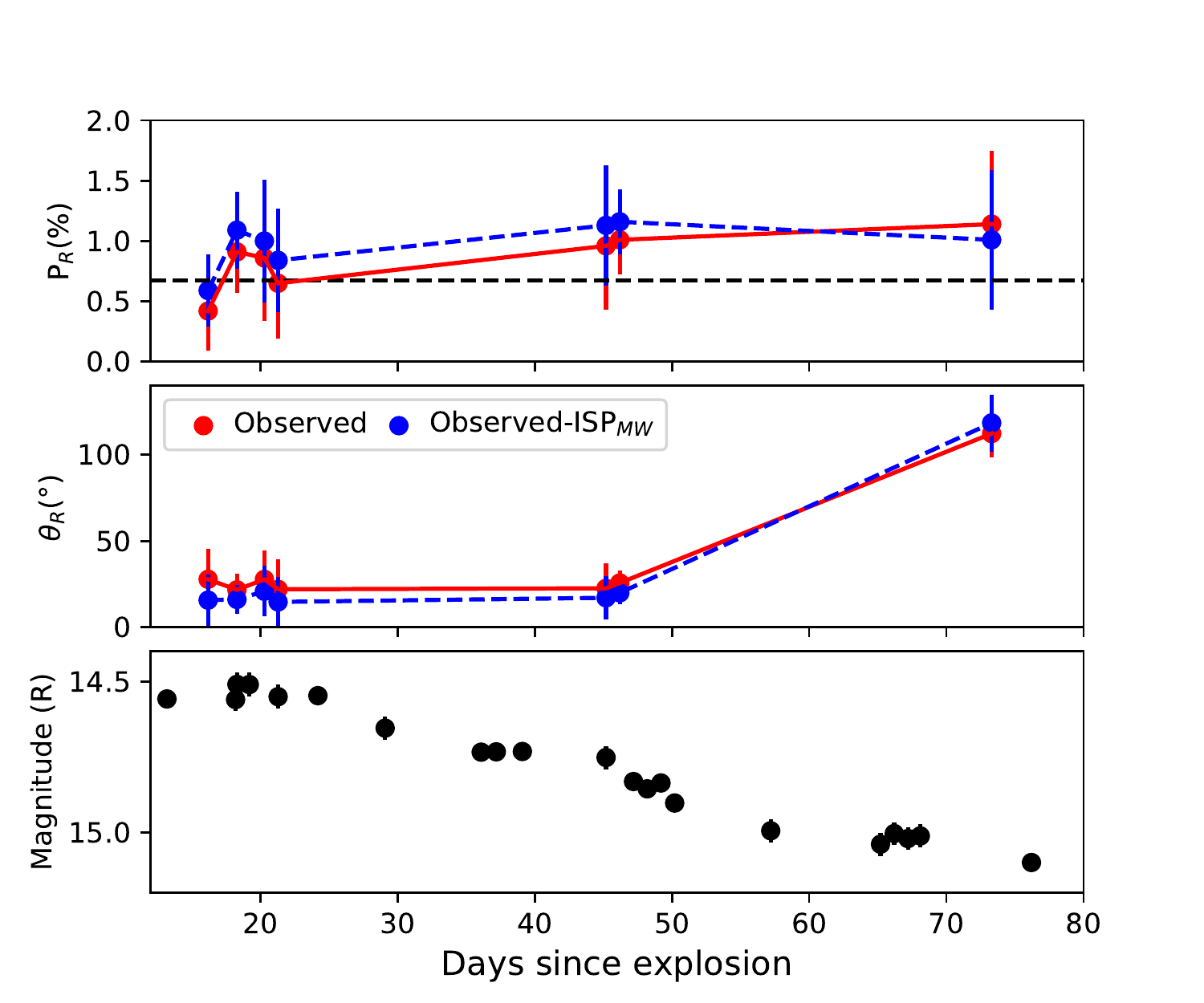}
	\caption{{\it Top panel:} Temporal evolution of degree of polarization in $R$-band (P$_R$) for SN~2016B. The horizontal dashed line represents the maximum  amount of host galaxy polarization. {\it Middle panel:} Temporal evolution of polarization angle $\theta$$_R$ for SN~2016B. The observed polarization parameters are shown in dashed lines and the ISP$_{MW}$ subtracted polarization parameters are shown in solid line. {\it Bottom panel}: $R$-band light curve of SN~2016B spanning the period of polarimetric observations.}
	\label{fig:pol_isp}
\end{figure}

\subsection{Estimation and evolution of intrinsic polarization} 
The intrinsic polarization of the SN is computed by vectorially subtracting the ISP$_{MW}$ from the observed SN polarization. The evolution of degree of polarization and the polarization angle during the photospheric phase of SN 2016B are presented in top and middle panel of Fig. \ref{fig:pol_isp} along with the $R$-band light curve evolution in this period in the bottom panel. The polarization parameters are tabulated in Table \ref{tab:pol_SN}. Since we have not subtracted the host galaxy polarization from the observed values, we caution the reader that these values are over- or underestimates of the intrinsic polarization (depending on the polarization angle of the host galaxy). While the ISP subtracted polarization remains nearly constant to 1~per~cent during the observation period from the early declining phase to the plateau phase, there is an abrupt change in the polarization angle at 73.5~d by $\sim$ 100$\degree$. The polarization angle in SN 2004dj has also been observed to change by 30$\degree$ during the fall from the plateau to the radioactive tail phase. In SN 2004dj, the possible reasons conjectured were the gradual unveiling of $^{56}$Ni in the inner regions distributed along the equatorial plane as the SN enters the nebular phase or likely an effect of opacity \citep{2006Natur.440..505L}. Nevertheless, in the absence of a good estimate of the interstellar polarization, suggesting a change in polarization angle is inconsequential due to its strong dependence on the degree of alignment of the ISP with the dominant axis of the SN ejecta. In the presence of a significant ISP, changes in the amount of polarization at fixed angle may be manifested as a change in polarization angle, arising the possibility of a change in the intrinsic polarization percentage in the final epoch and not just the angle. However, since we do not have another nearby epoch of polarimetric observation of SN~2016B, we defer from making any firm conclusion.

\begin{figure}
		\includegraphics[scale=1.00, width=0.48\textwidth,clip, trim={0.5cm 0.1cm 1.6cm 1.3cm}]{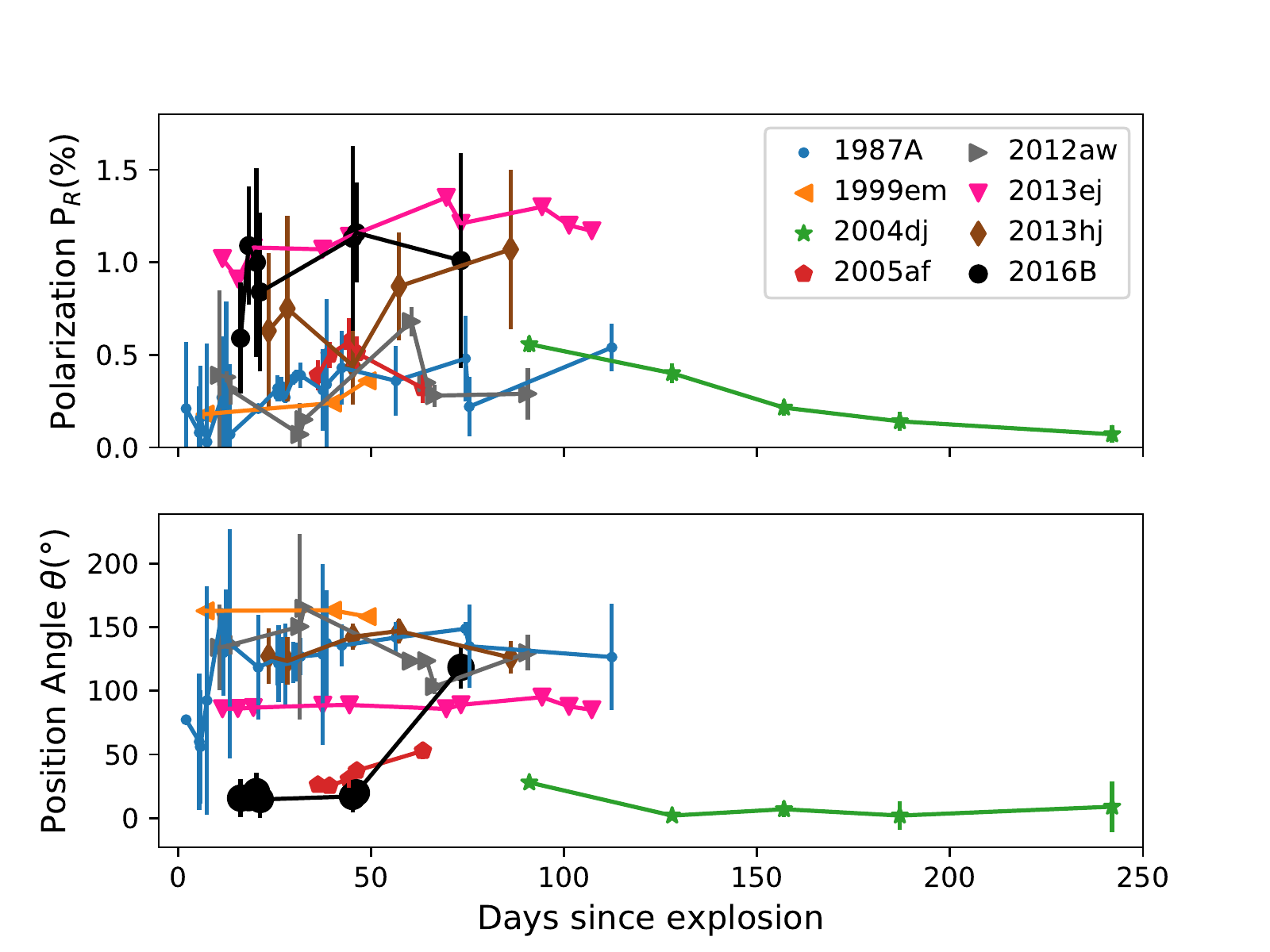}
	\caption{Temporal evolution of the ISP-subtracted polarization parameters of SN 2016B along with those of other Type II SNe.}
	\label{fig:pol_PA}
\end{figure}

In Fig. \ref{fig:pol_PA}, we compare the polarization parameters of SN~2016B with SNe~1987A \citep{1988MNRAS.234..937B}, 1999em \citep{2001ApJ...553..861L}, 2004dj \citep{2006Natur.440..505L}, 2005af \citep{2006A&A...454..827P}, 2012aw \citep{2014MNRAS.442....2K}, 2013ej \citep{2017ApJ...834..118M} and 2013hj \citep{2016MNRAS.455.2712B}. The polarization parameters of SN 1999em are obtained from $V$-band polarimetry, although, the evolution in $R$-band is expected to be similar. SNe~2013ej and 2016B exhibits the maximum polarization (\textgreater 1~per~cent) in the sample. The rest of the SNe except SN~2013hj show polarization values below 0.5~per~cent. However, considering the maximum host galaxy polarization, coaligned with the SN polarization, the intrinsic polarization of SN~2016B will drop down to $\sim$0.4~per~cent, similar to most SNe~II.

\begin{table*}
\caption{Polarimetric evolution of SN 2016B.}
\begin{tabular}{ccccccccc}
\hline
UT date & JD & Phase$^\dagger$ & \multicolumn{2}{c}{Observed} & \multicolumn{2}{c}{ISP$_{MW}$ subtracted} \\
(yyyy-mm-dd) & 245 7000+ & (d) &  P$_R$(\%) & $\theta_R$($\degree$) & P$_R$(\%) & $\theta_R$($\degree$) \\
\hline 
2016-01-10.9 & 398.4 & 16.4 & 0.42 $\pm$ 0.33 & 27.77 $\pm$ 17.72 & 0.59 $\pm$ 0.30 & 15.69 $\pm$ 14.87 &  &\\
2016-01-13.0 & 400.5 & 18.5 & 0.91 $\pm$ 0.34 & 21.66 $\pm$ 9.56 & 1.09 $\pm$ 0.32 & 16.09 $\pm$ 8.39 & &\\
2016-01-15.0 & 402.5 & 20.5 & 0.86 $\pm$ 0.52 & 27.77 $\pm$ 16.80 & 1.00 $\pm$ 0.51 & 20.84 $\pm$ 14.71 & &\\
2016-01-16.0 &  403.5 & 21.5 & 0.65 $\pm$ 0.46 & 22.00 $\pm$ 17.47 &  0.84 $\pm$ 0.43 & 14.66 $\pm$ 14.67 & & \\
2016-02-08.9 & 427.4 & 45.4 & 0.96 $\pm$ 0.53 & 22.56 $\pm$ 14.33 & 1.13 $\pm$ 0.50 & 17.06 $\pm$ 12.73 & & \\
2016-02-09.9 & 428.4 & 46.4 & 1.01 $\pm$ 0.29 & 25.66 $\pm$ 7.39 & 1.16 $\pm$ 0.27 & 19.92 $\pm$ 6.76 & & \\
2016-03-08.0 & 455.5 & 73.5 & 1.14 $\pm$ 0.61 & 112.32 $\pm$ 13.99 & 1.01 $\pm$ 0.58 & 118.48 $\pm$ 16.55 & & \\
\hline

\end{tabular}
\begin{tablenotes}
\item[a]$^\dagger$since explosion epoch t$_0$ = 2457382.0 JD (2015 December 25.4)
     \end{tablenotes}
\label{tab:pol_SN}   
\end{table*}

\section{Explosion parameters}
\label{sec7}
\subsection{Bolometric Light Curve}
\begin{figure}
	\begin{center}
		\includegraphics[scale=1.00, width=0.52\textwidth,clip, trim={0cm 0cm 0cm 1.3cm}]{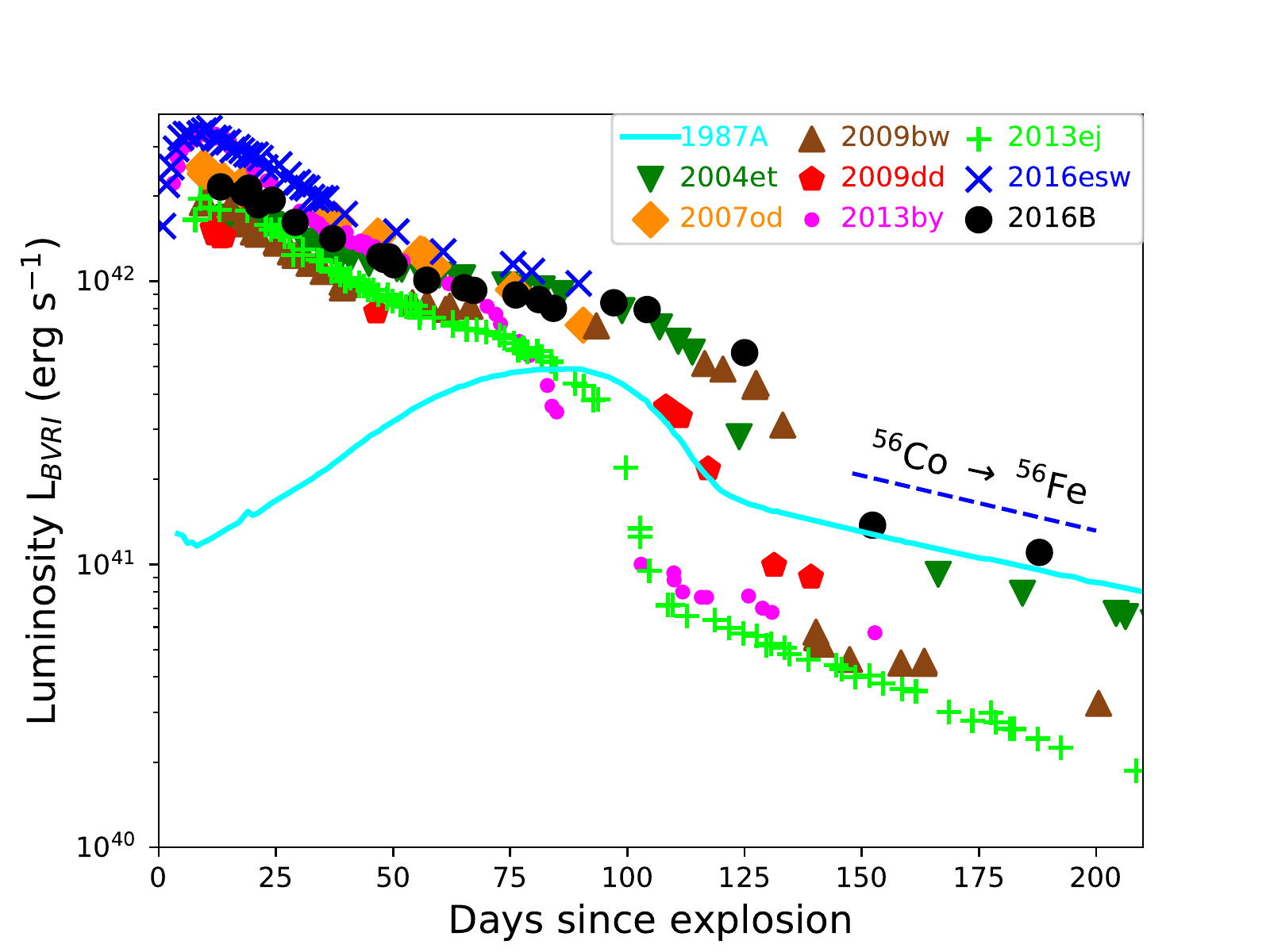}
	\end{center}
	\caption{The $BVRI$ pseudo-bolometric light curve of SN~2016B is compared with those of other SNe II from the comparison sample. The exponentially declining radioactive tail assuming full trapping of $\gamma$-ray photons and positrons is shown with dashed lines.}
	\label{fig:bol_light_curve}
\end{figure}

We compute the pseudo-bolometric luminosities of SN 2016B using $BVRI$ magnitudes following the method described in \cite{2018MNRAS.479.2421D}, which include SED integration using the trapezoidal rule over the photometric fluxes after correcting for extinction and distance. The zero points to convert the dereddened $BVRI$ magnitudes to fluxes are taken from \cite{1998A&A...333..231B}. Note that the $B$-band magnitude on 188.1~d is computed from the $(B-V)$ colour which becomes nearly constant at the late phases. We compared the pseudo-bolometric light curves of SN 2016B with those of other SNe from the comparison sample in Fig. \ref{fig:bol_light_curve}, constructed using the same method to maintain consistency. The shape of light curve of SN 2016B resembles that of SN 2007od. The evolution after the initial decline is similar to SNe 2004et and 2009bw, except that SN 2016B has a shorter effective plateau length and a more luminous radioactive tail compared to these two events, the latter indicating a greater amount of $^{56}$Ni synthesized. The tail phase luminosity is rather similar to SN 1987A. 

Further, to estimate the explosion parameters such as the $^{56}$Ni mass, the explosion energy and the ejecta mass, modelling of the true bolometric light curve is necessary. The true bolometric luminosities can be obtained by taking into account the contributions from the ultraviolet (UV) to the infrared (IR) bands. We construct the true bolometric light curve using the direct integration module in SuperBoL \citep{2017PASP..129d4202L} that is based on the method given in \cite{2009ApJ...701..200B}. This involves extrapolating the SED to the UV and IR regimes, assuming the SED to be blackbody. At later times, when the ejecta becomes optically thin and the SN can no longer be approximated by a blackbody, a linear extrapolation is performed in the UV wavelengths. We use the pseudo-bolometric luminosity to estimate the $^{56}$Ni mass in Section \ref{sect_ni} and the true bolometric luminosity is used in the analytical modelling as discussed in Section \ref{ana}.

\subsection{Mass of $^{56}$Ni}
\label{ni_mass}
The tail bolometric luminosity provides a tight constraint on the $^{56}$Ni mass synthesized during the explosion since the early tail phase is powered by the decay of $^{56}$Co, the daughter nucleus of $^{56}$Ni, to $^{56}$Fe. Below we use the tail bolometric luminosity to estimate the $^{56}$Ni mass in two ways.

\subsubsection{$^{56}$Ni mass from SN 1987A}
\label{sect_ni}
For SN 1987A, the mass of $^{56}$Ni has been determined fairly accurately to be 0.075$\pm$0.005~M$_\odot$ \citep{1996snih.book.....A}. Hence by assuming similar $\gamma$ ray energy deposition and comparing the bolometric luminosity with that of SN 1987A at a similar phase, we can estimate M$_{Ni}$ of SN 2016B. For comparison, we use the $BVRI$ bolometric luminosity at 188~d, which is (9.52$\pm$0.28)$\times$10$^{40}$ and (11.0$\pm$4.0)$\times$10$^{40}$~erg~s$^{-1}$ for SNe 1987A and SN 2016B respectively. The ratio of luminosities of SN 2016B and SN 1987A is 1.16$\pm$0.42, and hence the value of M$_{Ni}$ for SN 2016B is 0.087$\pm$0.032~M$_\odot$.

\subsubsection{$^{56}$Ni mass from tail luminosity}
\label{sect_tail}
The $^{56}$Ni mass is estimated from the tail bolometric luminosity using the expression given in \cite{2003ApJ...582..905H}:

\begin{equation}
 M_{Ni} = 7.866 \times 10^{-44} \times L_t exp\Bigg[\frac{(t-t_0)/(1+z) - 6.1}{111.26}\Bigg] M_\odot 
\end{equation}
where $t_0$ is the explosion time, 6.1 days is the half-life time of $^{56}$Ni and 111.26 days is the e-folding time of the $^{56}$Co decay. The tail luminosity ($L_t$) at $\sim$152.5 and 188.1~d is 2.60$\pm$1.06~$\times$~10$^{41}$ and 2.14$\pm$0.87 $\times$ 10$^{41}$~erg~s$^{-1}$ respectively, which corresponds to a mean $^{56}$Ni mass of 0.079 $\pm$ 0.023~M$_\odot$.

\begin{figure}
		\includegraphics[scale=1.00, width=0.53\textwidth,clip, trim={0cm 0cm 0cm 0cm}]{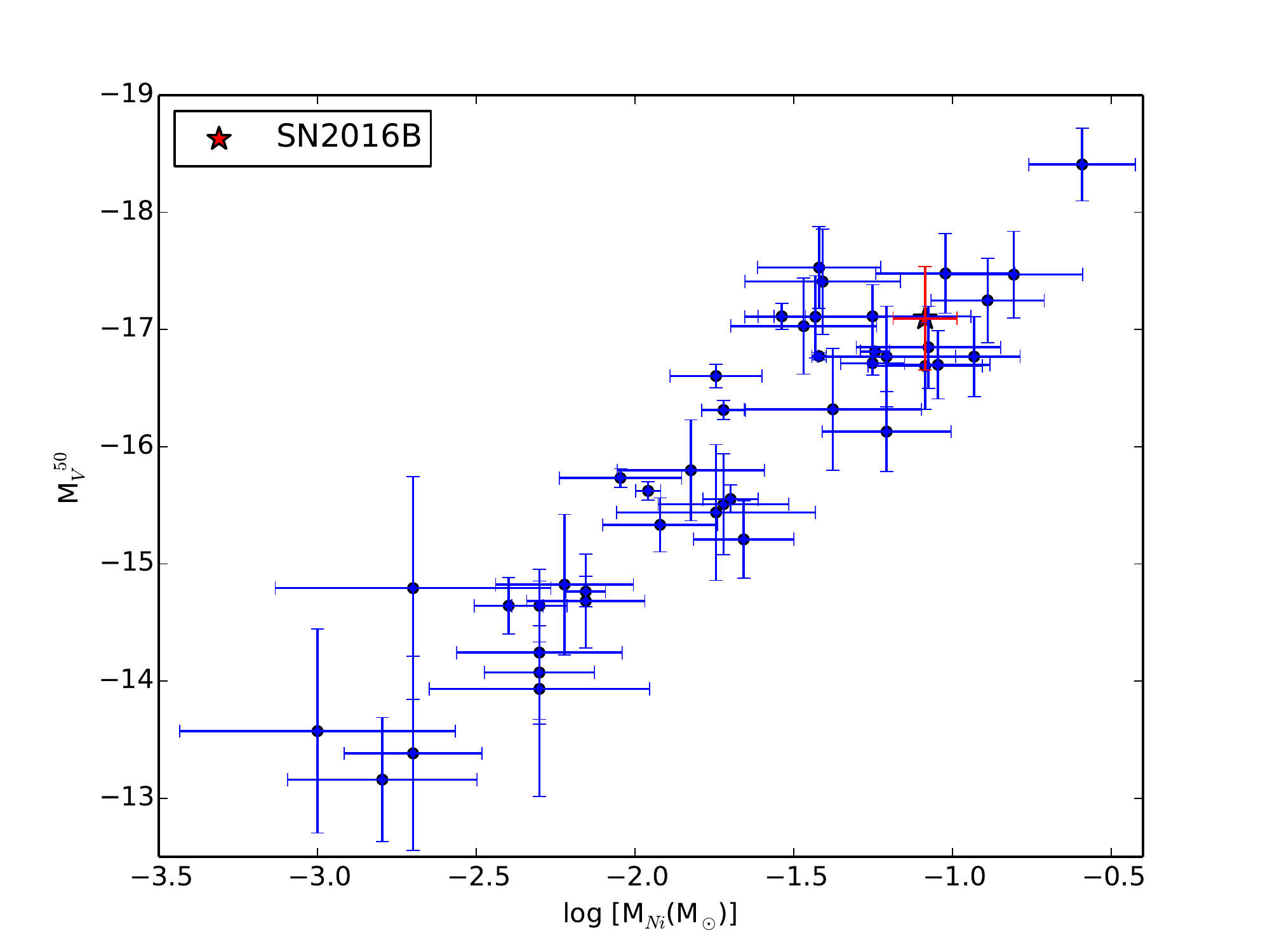}
	\caption{The position of SN~2016B in the absolute $V$-band magnitude at 50~d vs $^{56}$Ni mass is shown. The data is taken from \citet{2003ApJ...582..905H}, \citet{2014MNRAS.439.2873S}, and \citet{2015MNRAS.448.2608V}.}
	\label{fig:Mv_Mni}
\end{figure}

The above two methods yield consistent results and the weighted mean value of $^{56}$Ni mass from these methods is 0.082$\pm$0.019~M$_\odot$. SN 2016B, hence follows the mean absolute plateau magnitude at 50~d (M$_V^{50}$) and $^{56}$Ni mass (M$_{Ni}$) correlation as shown in Fig. \ref{fig:Mv_Mni}, which is reproduced using the data from \cite{2003ApJ...582..905H}, \cite{2014MNRAS.439.2873S} and \cite{2015MNRAS.448.2608V}.

\subsection{Semi-analytical Modelling}
\label{ana}
The comparison of the true bolometric light curve with those generated from semi-analytic models can provide reliable estimates of the physical parameters of the explosion. The two component semi-analytic model with a dense inner core and an extended low mass envelope developed by \cite{2016A&A...589A..53N} is used to generate the model light curves. The radiative diffusion model is used in this technique in order to incorporate the recombination energy to the total energy budget powering the light curve. This was originally developed by \cite{1989ApJ...340..396A} for a homologously expanding spherical ejecta, where the dominant source of opacity is assumed to be constant Thomson scattering opacity. The final bolometric luminosity is computed as the sum of the recombination energy and the energy liberated from radioactive decay of $^{56}$Ni and $^{56}$Co, taking the effect of gamma-ray leakage into consideration.

The expanding ejecta typically attains a constant density profile as it reaches the state of homologous expansion, while the outer part of the star develops a power law density profile. We used a constant density profile for the core and a power-law density profile with index (n) 2 for the shell. Similar density profile for the shell has been used to model the light curve of SN 2013ej, another transitional object (see \citealt{2016A&A...589A..53N} and \citealt{2017ApJ...838...28M}, for a discussion). The best fit ejecta mass, progenitor radius and explosion energy are 14 M$_\odot$, 2.8 $\times$ 10$^{13}$ cm ($\sim$ 400 R$_\odot$), and 1.4 foe respectively and the total ejecta mass, adding a shell mass of 0.53~M$_\odot$ obtained from the modelling, is 14.5~M$_\odot$. Assuming a core of 1.5~M$_\odot$, the pre-SN star has a mass of 16~M$_\odot$, which would mean that the zero-age main-sequence (ZAMS) mass of the progenitor lies in the top edge of the Type II progenitor mass distribution \citep{2015PASA...32...16S}. So far, the classical Type IIP SNe are found to arise from RSGs with ZAMS mass of \textless 15~M$_\odot$, while Type IIL-like events have slightly more massive yellowish progenitors \citep{2010ApJ...714L.254E,2010ApJ...714L.280F}. Thus, the progenitor of SN~2016B appears to be in the borderline between Type IIP and IIL-like progenitors. However, one should be careful while quoting the ejecta mass (M$_{ej}$) estimate from semi-analytic light curve models as the M$_{ej}$ and opacity ($\kappa$) are strongly correlated parameters, and hence introduces significant uncertainty in the determination of these parameters. It is only their product (M$_{ej}$.$\kappa$) which can constrained from the modelling. The mass of $^{56}$Ni estimated from the fit is 0.086 M$_\odot$, which is consistent with the value derived in Sect. \ref{ni_mass} within errors.

\begin{figure}
	\begin{center}
		\includegraphics[scale=1.00, width=0.56\textwidth,clip, scale=0.8, trim= {1.8cm 9.1cm 8cm 0cm}]{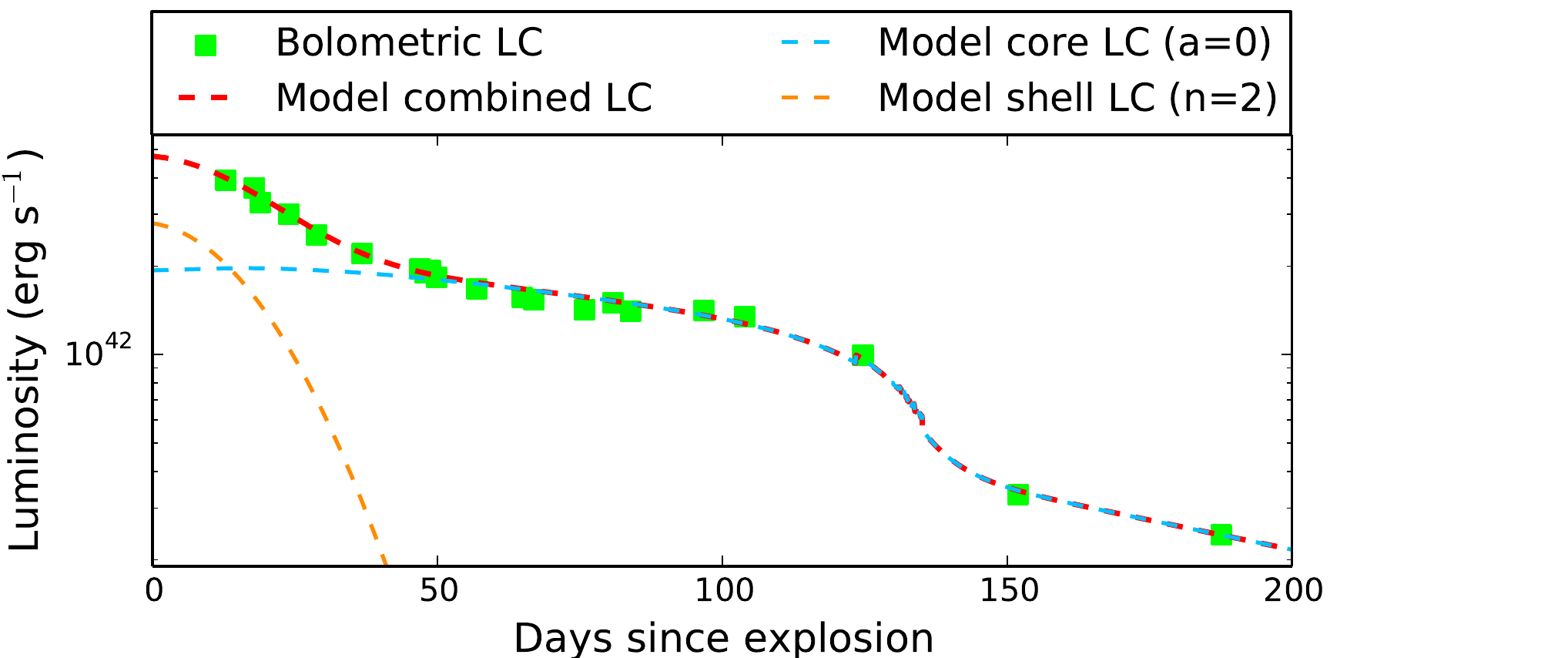}
	\end{center}
	\caption{The best fit two component analytical model of \citet{2016A&A...589A..53N} to the observed bolometric light curve of SN~2016B.}
	\label{fig:Nagy_Vinko}
\end{figure}

The parameters of the outer shell can also be estimated using this model, like the radius of the H-envelope (R = 11 $\times$ 10$^{13}$ cm). This can be considered as a lower limit, since dense sampling during the early phase is required for a reliable estimate. The true bolometric light curve with the best fit model is shown in Figure \ref{fig:Nagy_Vinko}. The parameters of the shell and core are listed in Table \ref{Nagy}. 

\begin{table}
\setlength{\tabcolsep}{2pt}
 \begin{minipage}{84mm}
  \caption{The best fit core and shell parameters for true bolometric light curve of SN 2016B using \citet{2016A&A...589A..53N}.}
  
  \begin{tabular}{@{}lccc@{}}
  \hline
  \hline
Parameter & Core & Shell & Remarks  \\
 \hline
R$_0$ (cm)  & 2.8 $\times$ 10$^{13}$ & 11 $\times$ 10$^{13}$ & Initial radius of ejecta\\
T$_{rec} (K)$ & 9500 & -  & Recombination temperature\\
M$_{ej}$ (M$_\odot$) & 14 & 0.53 & Ejecta mass\\
E$_{th} (foe)$ & 0.7 & 0.037 & Initial thermal energy \\
E$_{kin} (foe)$ & 0.7 & 0.073 & Initial kinetic energy \\
M$_{Ni}$ (M$_\odot$) & 0.086 & - & Initial $^{56}$Ni mass\\
$\kappa$ (cm$^2$ g$^{-1}$) & 0.15 & 0.2 &Opacity\\
A$_g$ (day$^2$) & 2.7 $\times$ 10$^{5}$ & 1 $\times$ 10$^{10}$ & Gamma-ray leakage \\
\hline
     \label{Nagy}
     \end{tabular}
\end{minipage}
\end{table}

\section{Spectral Analysis}
\label{sec8}
\begin{figure}
		\includegraphics[scale=1.00, width=0.5\textwidth,clip, trim={7.8cm 4.5cm 29.4cm 7.6cm}]{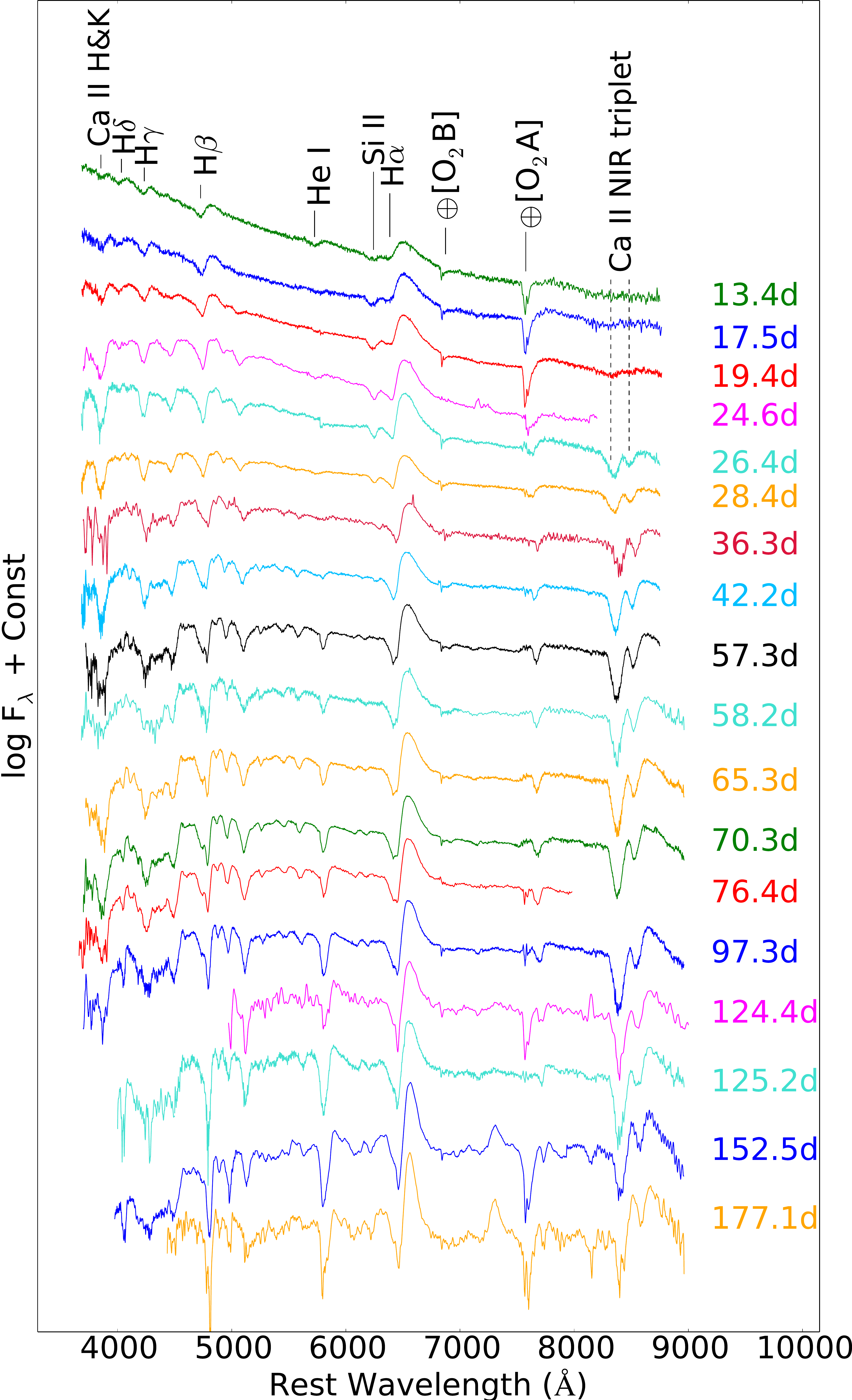}
	\caption{The spectral evolution of SN 2016B marked with the most prominent features in the spectra.}
	\label{fig:scaled_spectra}
\end{figure}

\begin{figure*}
		\includegraphics[scale=1.00, width=0.8\textwidth,clip, trim={1.1cm 0.35cm 1.5cm 1.5cm}]{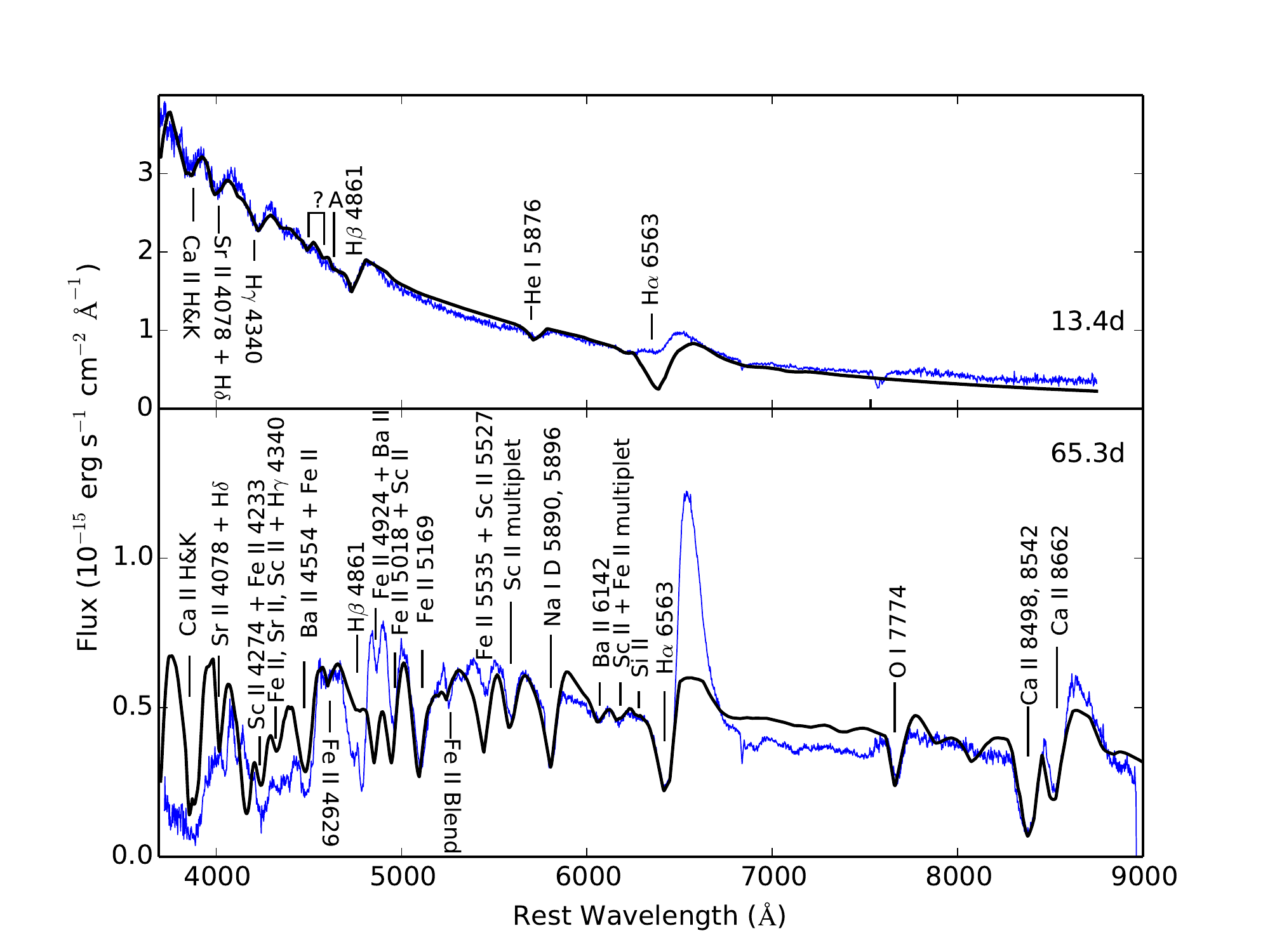}
	\caption{{\sc syn++} modelling of the early (13.4~d) and plateau phase (65.3~d) spectra of SN~2016B marked with the lines identified from the modelling.}
	\label{fig:syn}
\end{figure*}

The spectra of SN 2016B from 13.4 to 177.1~d from explosion is presented in chronological order in Fig. \ref{fig:scaled_spectra}. H~Balmer lines, the discerning feature of SNe II, are prominent throughout the evolution. Due to the multiple line blends, line identification in SN spectrum becomes difficult. We modelled the spectra of SN 2016B until the end of photospheric phase with {\sc syn++} \citep{2011PASP..123..237T} and a detailed line identification at two epochs, 13.4 and 65.3~d, which corresponds to the early and plateau phase of SN 2016B are shown in Fig. \ref{fig:syn}. {\sc syn++} is an updated version of {\sc synow} \citep{1997ApJ...481L..89F,1999MNRAS.304...67F,2002ApJ...566.1005B}, which uses the Sobolev approximation to simplify the radiative transport problem, and produce synthetic spectra, once the free expansion phase of the SN ejecta has ceased. It assumes the formation of spectral lines by resonance scattering above a sharp photosphere and a homologously expanding ejecta. 

\subsection{Salient Spectral Features}

The early spectra (13.4 - 19.4~d) are dominated by H {\sc i}, He {\sc i} and Ca {\sc ii} H\&K lines, which are fairly well-reproduced in the synthetic spectrum (Fig. \ref{fig:syn}). The proper fitting of the H$\delta$ model line in the observed spectra required the addition of Sr {\sc ii} ion in the modelling. An absorption dip bluewards of H$\alpha$ is seen and identified from the modelling as Si {\sc ii} $\lambda$6355 (Fig. \ref{fig:scaled_spectra}). The strength of this absorption feature is similar to H$\alpha$ at 13.4~d, becomes stronger than H$\alpha$ at 17.5 and 19.4~d, and then gradually weakens to disappear below the detection threshold at 57.3~d. From modelling, we tentatively identify a very weak feature bluewards of H$\beta$ (marked as \textquoteleft A\textquoteright{} in Fig. \ref{fig:syn}), as C {\sc ii} $\lambda$4745, which dies out after the 17.5~d spectrum. Two weak absorption dips at 4500 and 4583 \AA{} marked with \textquoteleft?\textquoteright{} in Fig.~\ref{fig:syn} are identified as normal and high velocity components of He {\sc ii} $\lambda$4686.

\begin{figure}
		\includegraphics[scale=1.00, width=0.48\textwidth,clip, trim={1.3cm 0.2cm 1.5cm 1.5cm}]{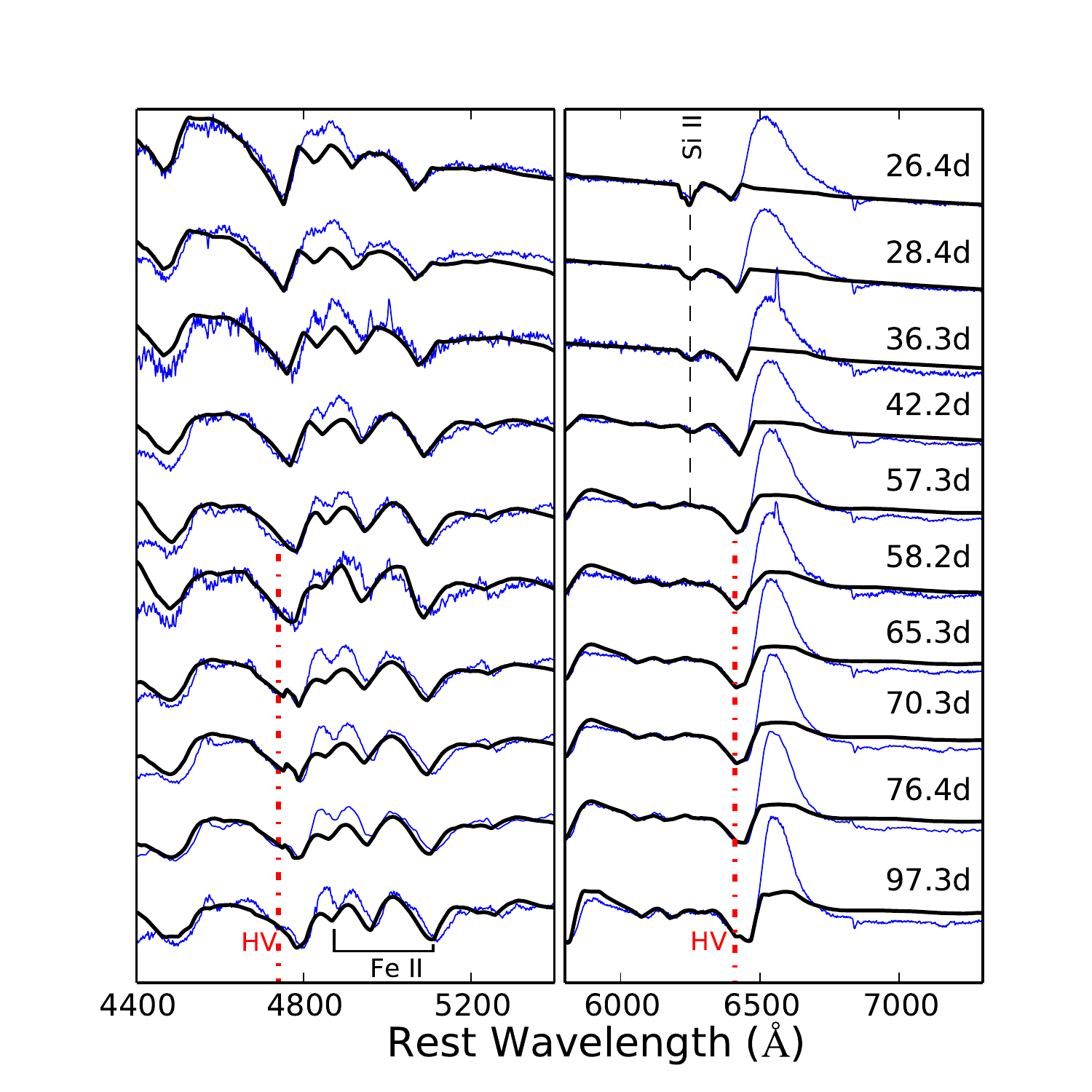}
	\caption{SYN++ modelling of the H Balmer, Fe {\sc ii} and Si {\sc ii} lines in the spectra of SN 2016B shown at the photospheric phases. The broad, extended absorption feature of H {\sc i} from the 57.3~d spectrum is reproduced using a normal velocity and a high velocity component.}
	\label{fig:model_Ha_Hb}
\end{figure}

\begin{figure}
		\includegraphics[scale=1.00, width=0.48\textwidth,clip, trim={2.0cm 0.2cm 2.0cm 1.0cm}]{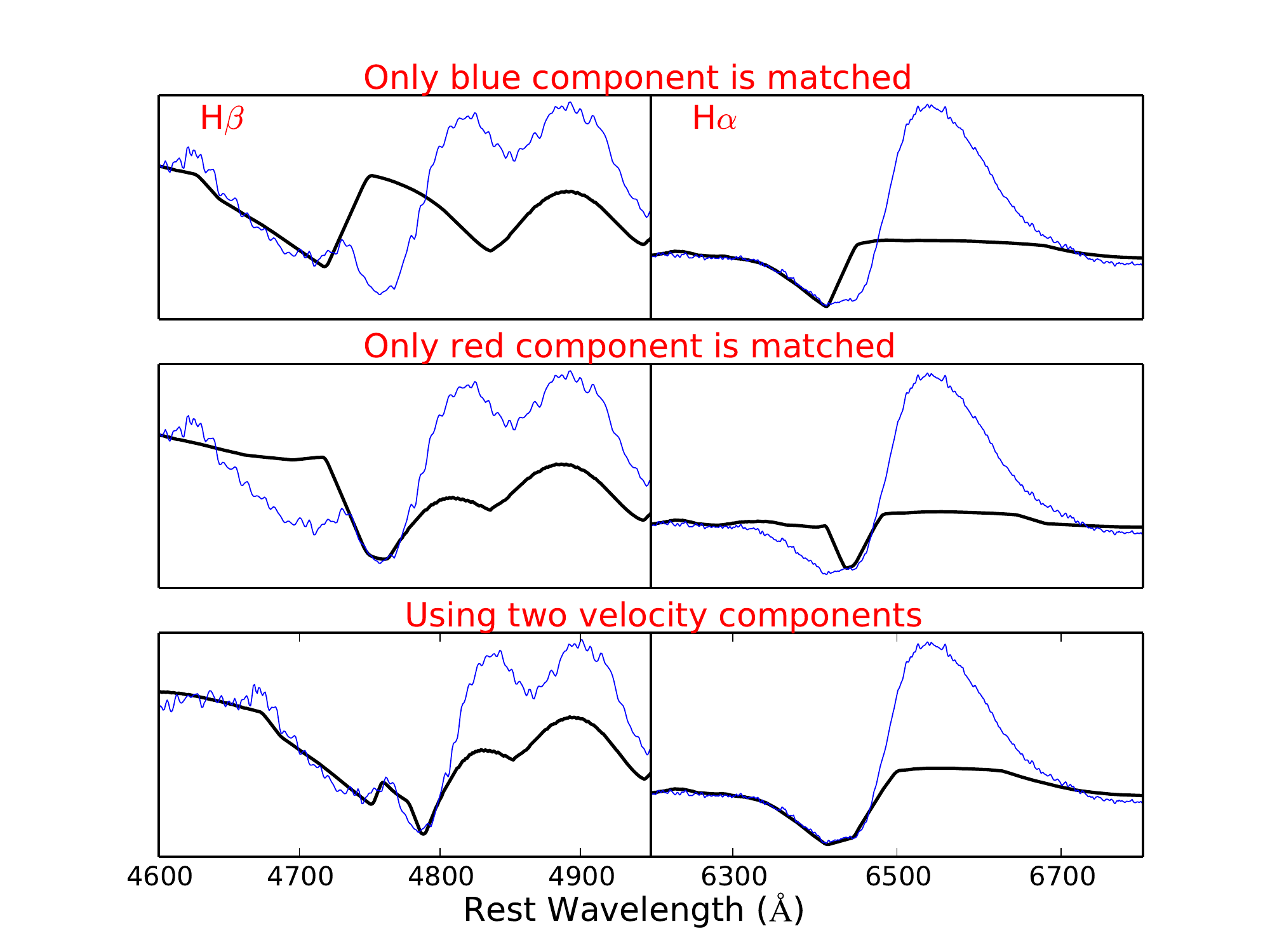}
	\caption{{\sc syn++} fits for the H Balmer lines in the 65.3~d spectrum, with the HV component in the upper panel, normal velocity component in the middle panel and both components in the bottom panel.}
	\label{fig:model_Ha_Hb_syn}
\end{figure}

The model fits to the H Balmer, Fe {\sc ii} and Si {\sc ii} lines in the observed photospheric phase spectra (26.4 - 97.3~d) are shown in Fig.~\ref{fig:model_Ha_Hb}, in which we mark the high velocity component of the H~{\sc i} lines conspicuous from the 57.3~d spectrum. In order to fit the extended H$\alpha$ and H$\beta$ profile, we used a combination of two velocity components i.e. variable normal velocity and constant high velocity (7000~km~s$^{-1}$) components, to reproduce H~{\sc i} lines in the 57.3 to 97.3~d spectra. In Fig. \ref{fig:model_Ha_Hb_syn}, we generate model fits for the H {\sc i} lines in the 65.3~d spectrum, with the low and high velocity (HV) component separately, to substantiate the point. The HV component may be arising from the excitation of the outer recombined ejecta by X-rays from the ejecta-CSM interaction in the photospheric phase \citep{2007ApJ...662.1136C}. Further, during the late photospheric phase, a notch forms due to excitation of the cool dense shell behind the reverse shock, which is apparent in the H$\beta$ feature in the spectra of SN~2016B. The stagnate nature of the HV component further strengthens the fact that this feature is not arising from the excitation of some metal line.

The early nebular phase spectra, at 152.5 and 177.1~d, are discerned by the appearance of forbidden lines of [O {\sc i}] $\lambda$6300, 6364 and [Ca {\sc ii}] $\lambda$7291, 7324. Further, we tentatively classify the feature at around 8130 \AA{} as a N~{\sc i} line blend. Other dominant elements in the nebular spectra are those with low quantum-level transitions, such as Na {\sc i} and Ca {\sc ii}.

\begin{figure}
	\begin{center}
		\includegraphics[scale=1.00, width=0.5\textwidth,clip, trim={1.2cm 0.3cm 1.6cm 1.2cm}]{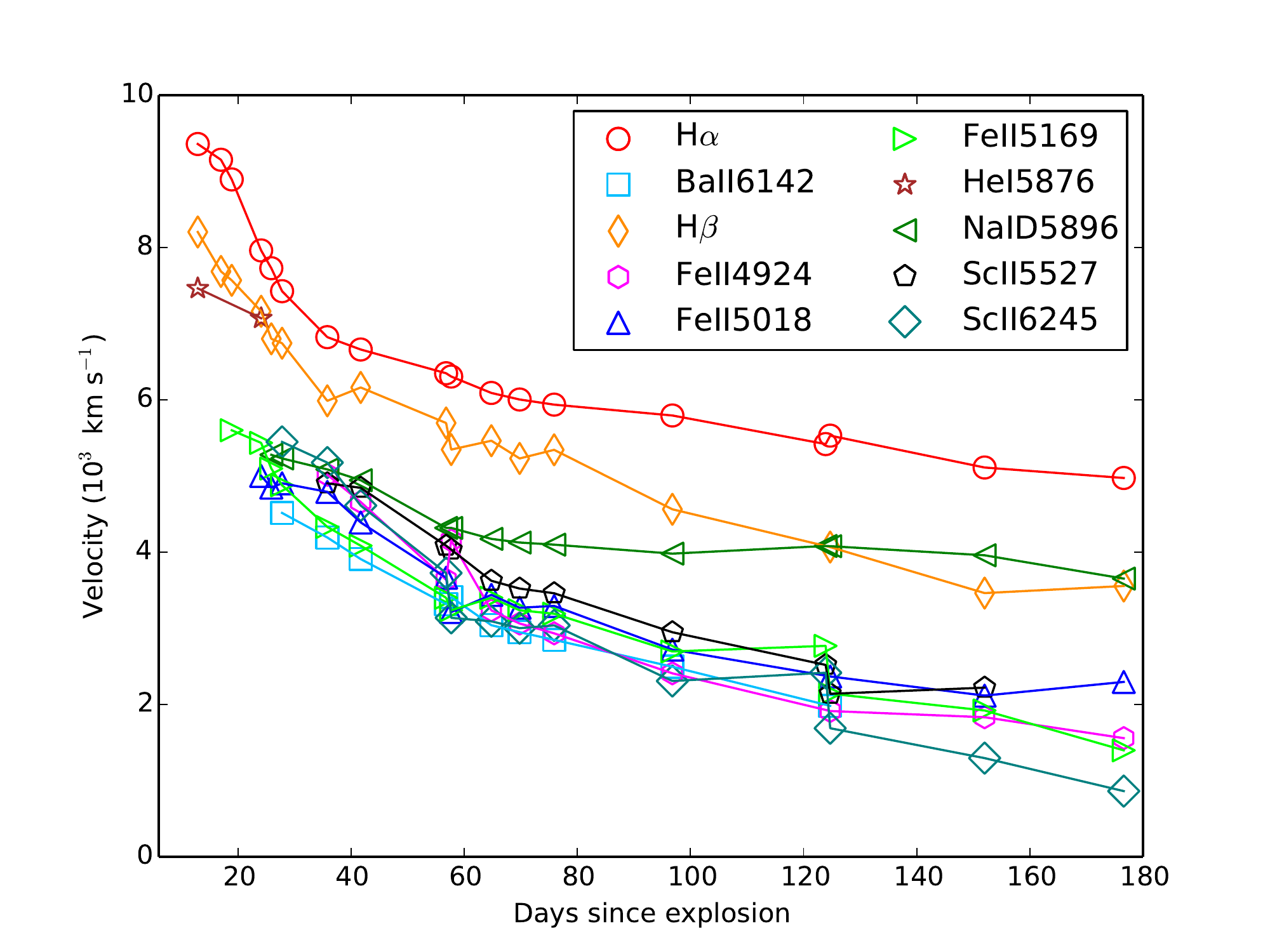}
	\end{center}
	\caption{Velocity evolution of the prominent lines in the spectra of SN 2016B estimated from the shift in the P-Cygni absorption feature.}
	\label{fig:vel_lines}
\end{figure}

In Fig. \ref{fig:vel_lines}, the evolution of the various prominent lines in the spectra of SN 2016B is shown. While the H$\alpha$ and H$\beta$ lines form at higher velocity, and hence at larger radii at all phases, the Fe {\sc ii}, Sc {\sc ii} and Ba {\sc ii} have nearly the same velocities and are formed at deeper layers of the ejecta.

\subsection{Comparison to other SNe II}
\begin{figure}
		\includegraphics[scale=1.00, width=0.5\textwidth,clip, trim={1.7cm 0.2cm 1.5cm 1.5cm}]{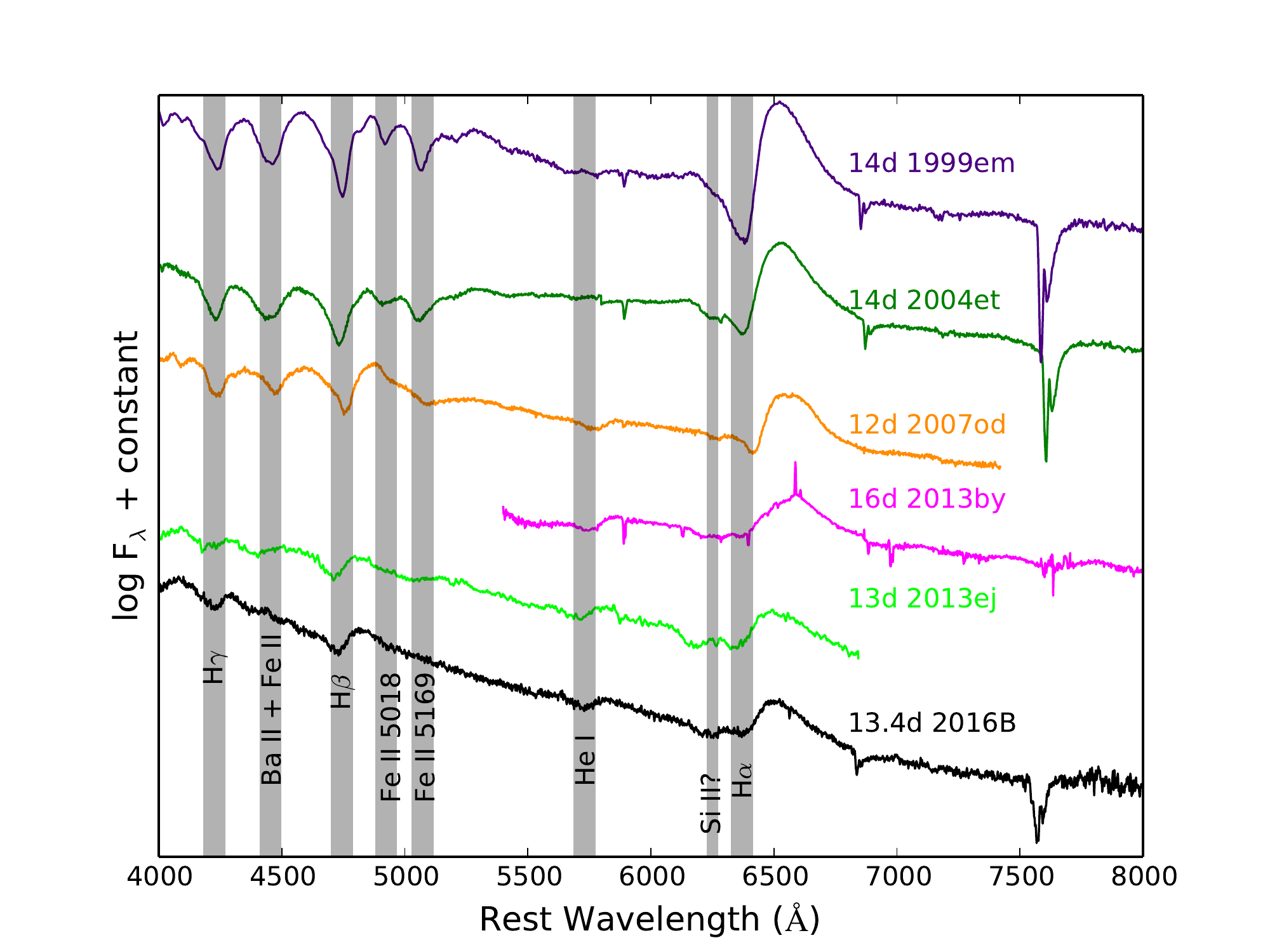}
	\caption{The early spectrum of SN 2016B at 13.4~d is compared with those of type IIP and IIL SNe.}
	\label{fig:spectra_comp_13d}
\end{figure}

A comparison plot of the 13.4~d spectrum of SN 2016B along with other SNe II is shown in Fig.~\ref{fig:spectra_comp_13d}. While the Fe {\sc ii} lines are quite well-developed in SNe 1999em and 2004et at coeval epochs, these lines have not yet emerged in SN 2016B, indicating higher ejecta temperature in SN~2016B. The absorption dip bluewards of H$\alpha$ in the spectra of SN 2016B, identified as Si {\sc ii} $\lambda$6355, is also conspicuous in the spectra of SNe 2004et, 2007od and 2013ej, although in SN 2004et this line is identified as the HV component of H$\alpha$ \citep{2006MNRAS.372.1315S}. 

\begin{figure}
		\includegraphics[scale=1.00, width=0.5\textwidth,clip, trim={1.7cm 0.2cm 1.5cm 1.5cm}]{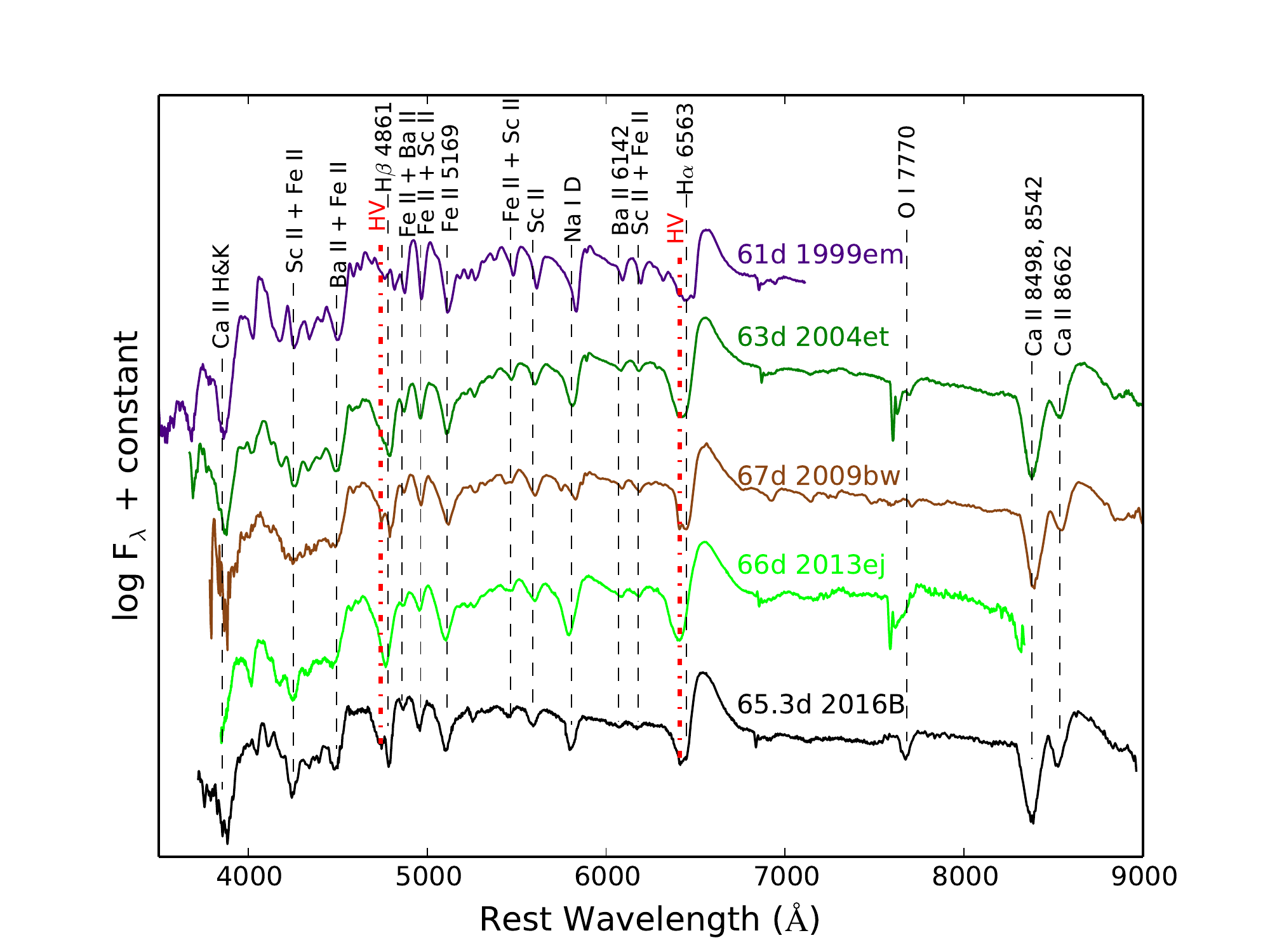}
	\caption{The plateau phase spectrum of SN 2016B at 65.3~d is compared with those of SNe II from the comparison sample.}
	\label{fig:spectra_comp_64d}
\end{figure}

\begin{figure}
		\includegraphics[scale=1.00, width=0.5\textwidth,clip, trim={1.6cm 0.2cm 1.5cm 1.5cm}]{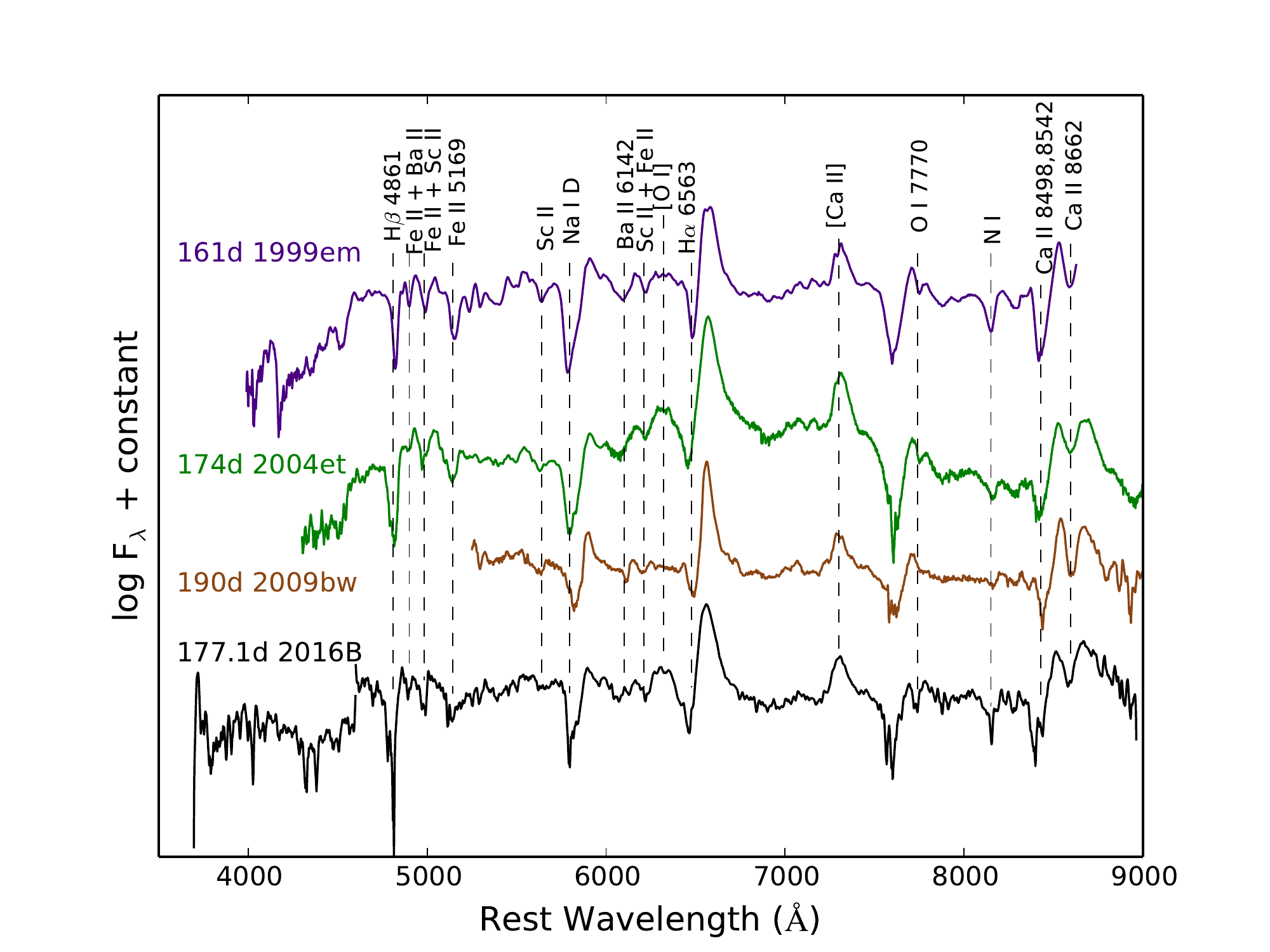}
	\caption{The comparison of the nebular phase spectrum of SN 2016B at 177.1~d with those of SNe II from the comparison sample.}
	\label{fig:spectra_comp_176d}
\end{figure}

The features in the plateau phase of SN 2016B is similar to those of other SNe II as shown in Fig.~\ref{fig:spectra_comp_64d}. HV component of H Balmer lines is prominent in SNe~2009bw, 2013ej and 2016B giving rise to broad absorption feature. In SN 2013ej, this HV component evolves with time \citep{2015ApJ...806..160B} and is non-evolving in SNe 2009bw \citep[7300~km~s$^{-1}$,][]{2012MNRAS.422.1122I} and 2016B (7000~km~s$^{-1}$). SNe~2013ej and 2016B display the strongest O~{\sc i}~$\lambda$7774 feature at this phase, while SN~2009bw show no evidence of oxygen. 

Due to the decreasing velocity and low opacity of ejecta at the nebular phase, the P-Cygni absorption trough of H {\sc i} and metal lines becomes narrower and weaker. The N~{\sc i} line blend identified in the nebular spectra of SN 2016B at 8130 \AA{} is also conspicuous in the spectra of SNe 1999em, 2004et and 2009bw (Fig.~\ref{fig:spectra_comp_176d}). 

\subsection{Progenitor mass from nebular spectra}
\begin{figure}
		\includegraphics[scale=1.00, width=0.5\textwidth,clip, trim={1.3cm 0cm 1.5cm 1.0cm}]{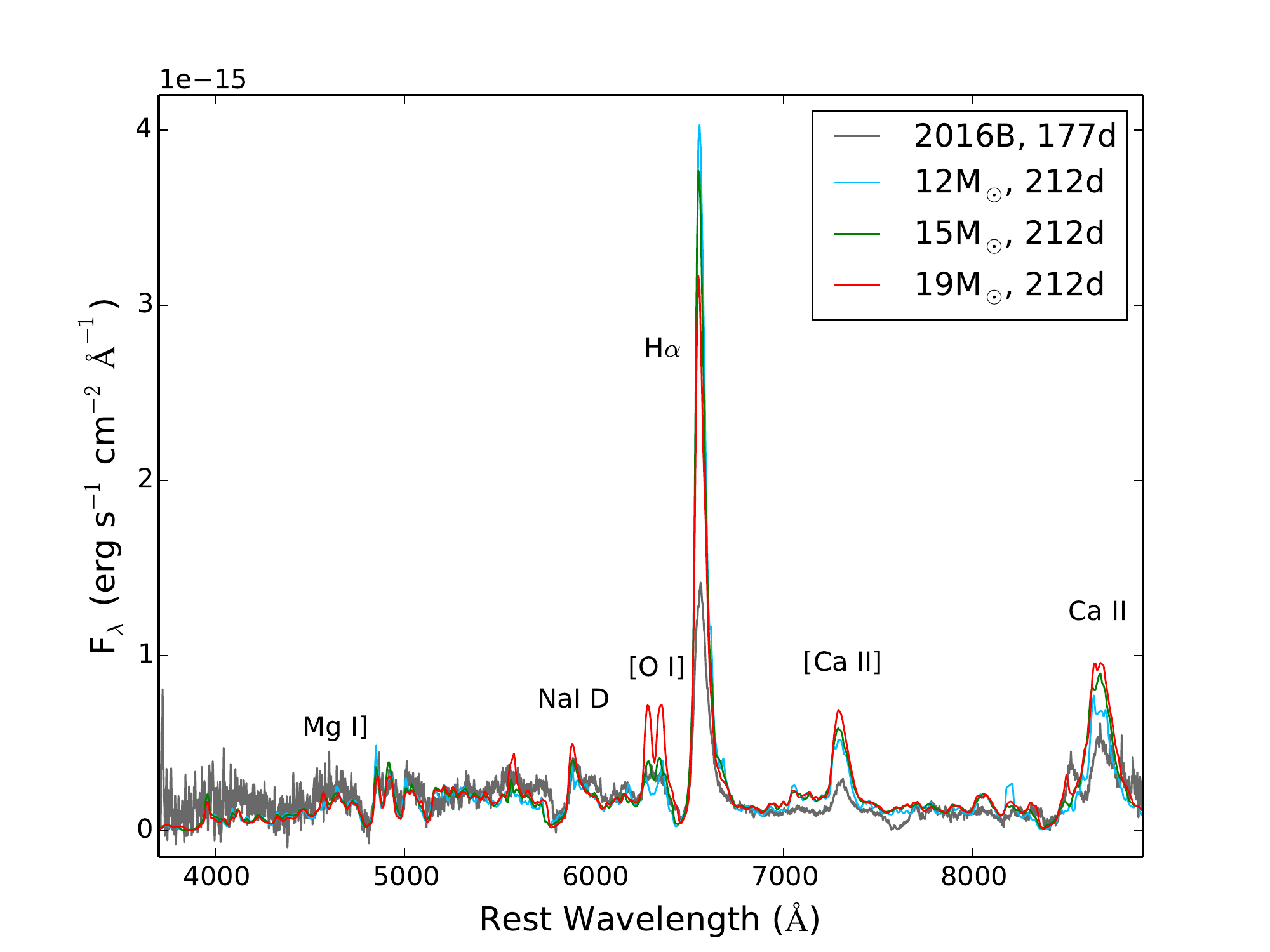}
	\caption{Comparison of the 177~d spectrum of SN 2016B with that of the scaled model spectra from \citet{2014MNRAS.439.3694J}.}
	\label{jerk_comp}
\end{figure}

As suggested by \cite{1995ApJS..101..181W}, the nucleosynthetic yield depends strongly on the mass of progenitor, hence the observed strengths of nebular lines can serve as a diagnostic of the progenitor mass. \cite{2014MNRAS.439.3694J} generated models of nebular spectra for 12, 15, 19 and 25 M$_\odot$ progenitor for 0.062 M$_\odot$ of $^{56}$Ni, assuming a distance of 5.5 Mpc. In Fig. \ref{jerk_comp}, we compare the 177~d spectrum of SN 2016B to these models to get a handle on the mass of progenitor. For the comparison, we scaled the model spectra with the ratio (0.082/0.062) to correct for $^{56}$Ni mass and then rescale the 212~d model flux to the distance and phase (177~d) of the nebular spectrum of SN 2016B. The SN spectrum is de-redshifted, dereddened and corrected for flux losses using the photometric fluxes, before comparison.

We find that the model spectra reproduces strong H Balmer and Ca {\sc ii} NIR triplet features. The strong H~{\sc i} line could be arising from the uncertainty in mixing scheme used for bringing hydrogen clumps from the outer layers into the radioactive core. 
The overestimated peak strength of Ca {\sc ii} NIR triplet is possibly due to a high density envelope used in the model, scattering the Ca {\sc ii} $\lambda$8498 and Ca {\sc ii} $\lambda$8542 in Ca {\sc ii} $\lambda$8662. The strength of [O {\sc i}] $\lambda\lambda$6300, 6364 in the 177~d spectrum of SN 2016B is more similar to that of the 15~M$_\odot$ (ZAMS) model, while the 12 and 19 M$_\odot$ models under- and overestimate the strength of the line, respectively. Also, the strength of Na {\sc i} D line in the spectrum of SN~2016B matches roughly with that of the 15~M$_\odot$ model.

Oxygen is the most abundant element in the carbon ashes (the O/Ne/Mg layer) and hence a large fraction of the thermalized energy in this layer is reemitted by oxygen in the form of the nebular line [O {\sc i}] $\lambda\lambda$6300, 6364. Thus, the [O {\sc i}] luminosity is a key to probe the amount of nucleosynthesized oxygen, which being highly sensitive to the helium core mass, is an indicator of the ZAMS mass (e.g. \citealt{1995ApJS..101..181W,1996ApJ...460..408T}, however, see \citealt{2016ApJ...818..124E} for a different perspective). To better constrain the progenitor mass, we estimate the [O~{\sc i}] luminosity for SN 2016B in the 177~d spectrum and compare to that of the model. The 212~d [O {\sc i}] luminosity for the model spectra are 1.36, 2.08 and 3.19 $\times$ 10$^{39}$ erg~s$^{-1}$ for 12, 15 and 19 M$_\odot$ progenitor mass, respectively. After scaling the luminosities for $^{56}$Ni mass and flux decline rate, the [O~{\sc i}] luminosities at 177~d are 2.00, 3.06, and 4.02 $\times$ 10$^{39}$ erg~s$^{-1}$ for 12, 15 and 19 M$_\odot$ model spectra, respectively. The [O~{\sc i}] luminosity for SN~2016B in the 177~d spectrum, obtained in a similar way as described in \cite{2012A&A...546A..28J}, is (2.66$\pm$0.97) $\times$ 10$^{39}$ erg~s$^{-1}$, which indicates that the ZAMS mass of the progenitor lies in the range of 12-17 M$_\odot$, taking into account the error in luminosity. The derived ZAMS progenitor mass range includes the pre-supernova star mass estimated from analytical modelling of the light curve in Section \ref{ana}.

\section{SN~2016B: A Type IIP or IIL SN?}
\label{sec9}
SN~2016B displays photometric and spectroscopic properties intermediate to those of Type~IIP and IIL SNe.
The post-peak decline rate in $V$-band is steeper (0.84~mag~50~d$^{-1}$) than that suggested by \cite{2014MNRAS.445..554F} to classify a SN as IIP and lasts up to $\sim$66~d, while the plateau decline rate is shallow (0.24~mag~50~d$^{-1}$) extending up to $\sim$118~d, and falls within the range of plateau slopes of SNe~IIP. The steep drop from the plateau to the radioactive tail, a feature considered for long to be the signature of Type IIP SNe, is 1.6~$\pm$~0.2~mag in 25~d in SN~2016B, in accord with the range of values in SNe II \citep[1.0 - 2.6 mag,][]{2016MNRAS.459.3939V}. However, this property has been noted in Type~IIL SNe observed for a longer duration of time \citep{2015MNRAS.448.2608V}. Further, the long optically thick phase duration of 118~d of SN~2016B, concurs with that of Type IIP SNe.

\begin{figure*}
		\includegraphics[scale=1.00, width=1.0\textwidth,clip, trim={3.0cm 0cm 3.4cm 1.3cm}]{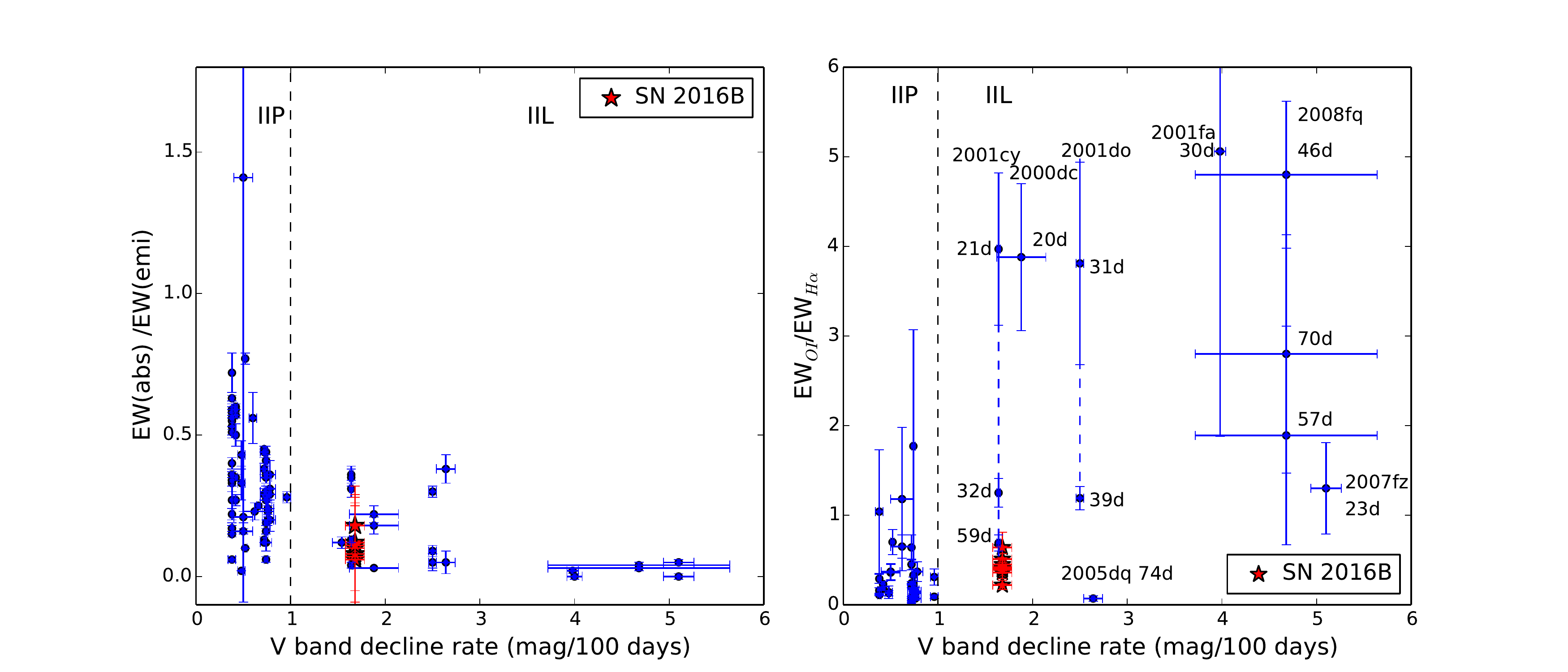}
	\caption{EW ratio of the absorption to emission feature of H$\alpha$ (left panel) and EW ratio of O~{\sc i}~$\lambda$7774 and H$\alpha$ (right panel) at different phases (26.4, 36.3, 42.2, 57.3, 58.2, 65.3, 70.3, 76.4, 97.3 and 125.2~d) in the spectra of SN 2016B (shown with red stars) compared with those in the \citet{2014MNRAS.445..554F} sample of SN II. The dashed line demarcates the IIP/IIL regimes based on the early decline rate as suggested in \citet{2014MNRAS.445..554F}.}
	\label{fig:Faran}
\end{figure*}

\begin{figure}
	\begin{center}
		\includegraphics[scale=1.00, width=0.48\textwidth,clip, trim={0.8cm 1.75cm 1.6cm 1.3cm}]{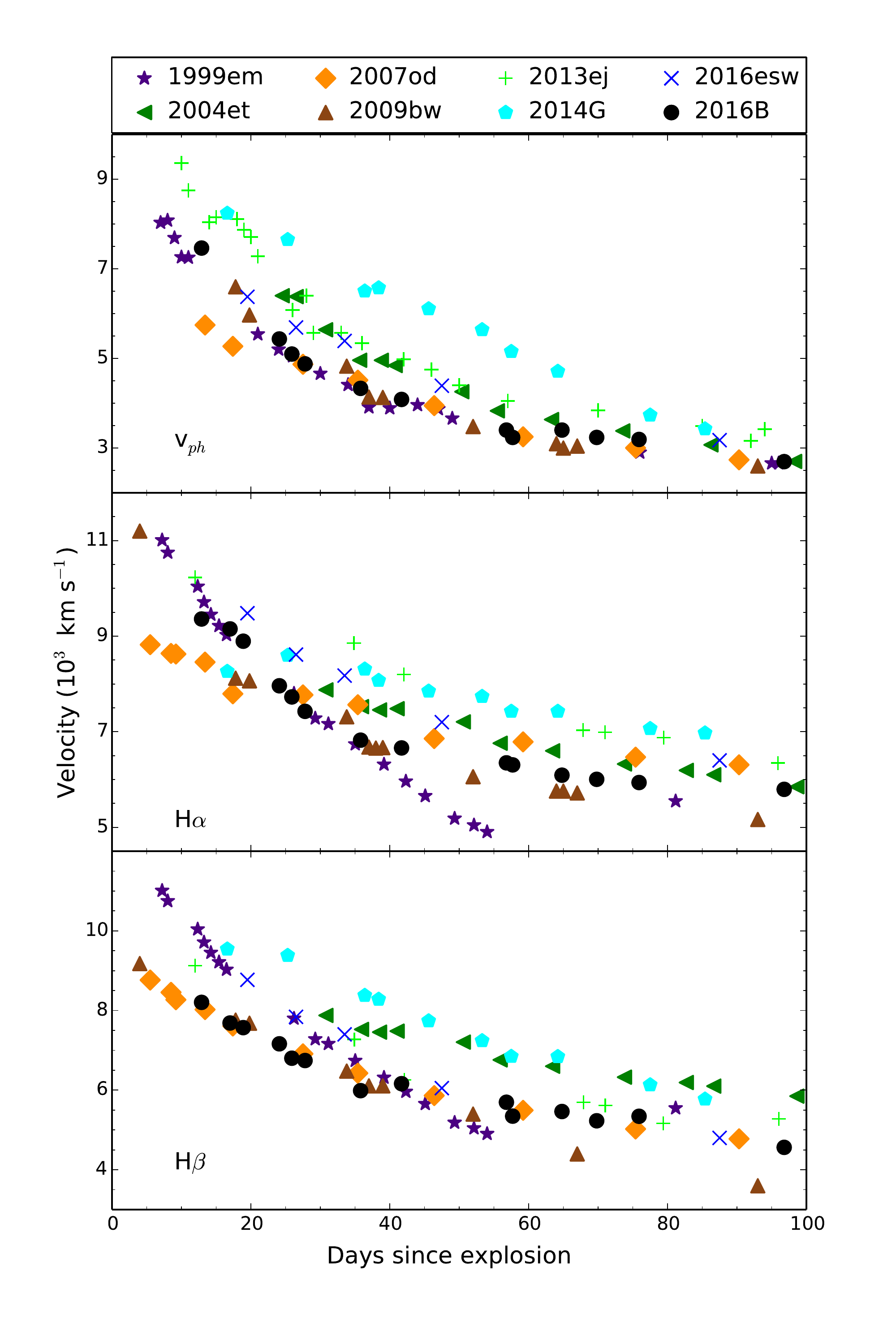}
	\end{center}
	\caption{The comparison of photospheric (v$_{ph}$; from the absorption minima of He~{\sc i} at early phases and Fe~{\sc ii}~5169 at photospheric stage), H$\alpha$ and H$\beta$ velocity evolution of SN 2016B with those of other Type II SNe.}
	\label{fig:vel_comp}
\end{figure}

While it is challenging to judge from spectroscopy whether an event is to be categorised as Type IIP or IIL, \cite{1994A&A...282..731P} and \cite{1996AJ....111.1660S} found from their analysis that SNe~IIL have shallower H$\alpha$ P-Cygni profile than SNe IIP, which is reinforced in the studies of \cite{2014ApJ...786L..15G} and \cite{2014MNRAS.445..554F}. This is possibly linked with the mass and density profile of hydrogen envelope in the pre-supernova star. Further, the metal lines (such as, O {\sc i} $\lambda$7774) in the spectra of SNe IIL are found to be stronger than that in SNe~IIP, which results from the rapid recombination wave in SNe IIL, leading to a highly dense photosphere. Thus, in order to discern the collocation of SN~2016B in the SN II zoo from its spectral properties, we reproduce the absorption to emission ratio of H$\alpha$ (a/e) vs $V$-band decline rate and EW$_{OI}$/EW$_{H\alpha}$ vs $V$-band decline rate plots from \cite{2014MNRAS.445..554F} in Fig.~\ref{fig:Faran}. In the former, we find that SN 2016B belongs to Type IIL class, and in the latter, the O {\sc i} line is weaker than in most Type~IIL SNe, and is more similar to Type~IIP SNe. Besides, the HV feature discernable in the H~{\sc i} lines at the photospheric phase, endorses SN~2016B as a Type~IIL SN.

Moreover, \cite{2014MNRAS.445..554F} suggested a higher photospheric velocity and shallower H~{\sc i} velocity evolution profiles of SNe IIL with respect to SNe IIP. The flatter velocity profile of H~{\sc i} in SNe IIL is attributed to the formation of absorption troughs in the outer, higher velocity layers above the photosphere, as the photosphere recedes into the hydrogen poor layers during the photospheric stage. We compare the photospheric velocity of SN~2016B (computed from the blue-shifted absorption trough of Fe {\sc ii} and Sc {\sc ii} lines at photospheric phase and He {\sc i} line at early phase) with those of other SNe~II in the top panel of Fig.~\ref{fig:vel_comp}. The photospheric velocity of SN~2016B is similar to SNe 1999em, 2007od and 2009bw, whereas, the velocity is lower by about 500-600~km~s$^{-1}$ than SNe 2013ej, 2004et and 2016esw. The middle and bottom panel of Fig.~\ref{fig:vel_comp} shows the velocity evolution of H$\alpha$ and H$\beta$ lines, respectively. In both the plots, SNe 1999em and 2009bw exhibits a power law decline, typical of Type IIP, while the flatter evolution of H~{\sc i} line velocity is apparent in SNe 2004et \citep{2006MNRAS.372.1315S}, 2007od \citep{2011MNRAS.417..261I}, 2013ej \citep{2015ApJ...806..160B}, 2014G \citep{2016MNRAS.462..137T}, 2016esw \citep{2018MNRAS.478.3776D} and 2016B, typical of Type IIL. 

The progenitors of SNe~II detected in pre-explosion $HST$ images and reported in \cite{2015PASA...32...16S} are found to have masses below 15~M$_\odot$ for most SNe IIP and are slightly more massive, yellow supergiants for SNe IIL. Further, we estimate the progenitor mass of SN~2016B by comparing the nebular spectrum to late-time spectral models. These methods yield values corresponding to the upper end of the range of masses of progenitors detected for SNe II (16.5$\pm$1.5~M$_\odot$), which commonly give rise to SNe IIL. From the above discussion, we conclude that SN~2016B exhibits intermediate properties in the SN II diversity.

\section{Conclusions}
\label{sec10}
In this paper, we present the analysis of the photometric, polarimetric and spectroscopic observations of SN~2016B in the galaxy PGC~037392. The post-peak decline in SN~2016B is steeper than most SNe classified as Type IIP in the literature. However, the optically thick phase duration is longer than SNe IIL lasting up to 118~d. In addition, the maximum recorded luminosity in $V$-band in the present study and the plateau decline rate does not corroborate with the correlation deduced from a sample of SNe II by \cite{2014ApJ...786...67A}. The plateau decline rate of SN 2016B quoted in a recent paper by \cite{2018MNRAS.479.3232G} as 0.34~$\pm$~0.01~mag~100~d$^{-1}$, which is within errors of our derived value (0.47~$\pm$~0.24~mag~100~d$^{-1}$). The low cadence photometric follow-up during the recombination phase is possibly the cause of the high error in our quoted light curve parameters, however, we could determine all the parameters reasonably well. 

We also report $R$-band polarimetric observations of SN~2016B. SN~2016B displays significant, but nearly constant intrinsic polarization throughout our observation period spanning from the early cooling to the recombination phase. Though the polarization angle shows an abrupt change at 73.5~d, we do not step into a conclusion as we do not have observation of any nearby epoch to support this value. The polarization of SN~2016B, considering the maximum host galaxy polarization coaligned with the SN polarization, is about 0.3$-$0.4~per~cent, which is typical of hydrogen rich Type II SNe in the photospheric phase.

We derived the $^{56}$Ni mass from the tail luminosity to be 0.082~$\pm$~0.019~M$_\odot$. Thus, SN~2016B falls within the plateau luminosity vs $^{56}$Ni mass correlation for SNe II \citep{2003ApJ...582..905H,2014MNRAS.439.2873S,2015MNRAS.448.2608V}. The explosion parameters derived from analytical modelling of the true bolometric light curve indicates a total ejecta mass of $\sim$15~M$_\odot$, an initial radius of 400~R$_\odot$ and a total explosion energy of 1.4~foe. The progenitor mass derived by comparing the nebular spectrum of SN~2016B to the spectral models from \cite{2014MNRAS.439.3694J}, suggests a 12-17~M$_\odot$ ZAMS progenitor. 

Broad absorption troughs of H {\sc i} lines is apparent from the 57.3~d spectrum, which we reproduced with a combination of normal velocity and constant HV component in {\sc syn++} modelling. HV component of H {\sc i} lines indicate a possible role of circumstellar interaction, more evident in Type IIL SNe. Moreover, the spectra of SN~2016B exhibits weaker absorption profile of H~{\sc i} lines similar to Type IIL SNe whereas weaker metal lines such as O~{\sc i}~7774 line unlike Type IIL SNe. The transitional nature of SN~2016B showing resemblance to both the photometric and spectroscopic properties of Type IIP and IIL, blurs out the fine line demarcating these two types, suggesting a continuum of properties in the hydrogen-rich SNe~II family.

\section*{Acknowledgments}
This work is partially based on observations collected at Copernico 1.82m telescope (Asiago, Italy) of the INAF - Osservatorio Astronomico di Padova. We thank N. Elias-Rosa for the NOT observations. This work made use of The Neil Gehrels $Swift$/UVOT data reduced by P. J. Brown and released in the $Swift$ Optical/Ultraviolet Supernova Archive (SOUSA). SOUSA is supported by NASA's Astrophysics Data Analysis Program through grant NNX13AF35G. This research has made use of the NASA/IPAC Extragalactic Database (NED) which is operated by the Jet Propulsion Laboratory, California Institute of Technology, under contract with the National Aeronautics and Space Administration. We acknowledge the usage of the HyperLeda data base (http://leda.univ-lyon1.fr). This work has made use of data from the European Space Agency (ESA) mission {\it Gaia} (\url{https://www.cosmos.esa.int/gaia}), processed by the {\it Gaia} Data Processing and Analysis Consortium (DPAC,
\url{https://www.cosmos.esa.int/web/gaia/dpac/consortium}). Funding for the DPAC has been provided by national institutions, in particular the institutions participating in the {\it Gaia} Multilateral Agreement.  SBP and KM acknowledges BRICS grant DST/IMRCD/BRICS/Pilotcall/ProFCheap/2017(G) for the present work. KM acknowledges the support from the IUSSTF WISTEMM fellowship and UC Davis. AP, LT, SB are partially supported by the PRIN-INAF 2016 with the project \textquotedblleft Towards the SKA and CTA era: discovery, localisation, and physics of transient sources\textquotedblright{} (P.I. M. Giroletti). BK acknowledges the Science and Engineering Research Board (SERB) under the Department of Science \& Technology, Govt. of India, for financial assistance in the form of National Post-doctoral Fellowship (Ref. no. PDF/2016/001563).

\bibliographystyle{mnras}
\bibliography{refag}   



\appendix

\section{Photometry}
\label{phot}
\begin{figure}
	\begin{center}
		\includegraphics[scale=1.00, width=0.5\textwidth,clip, trim={0.0cm 0cm 0cm 0cm}]{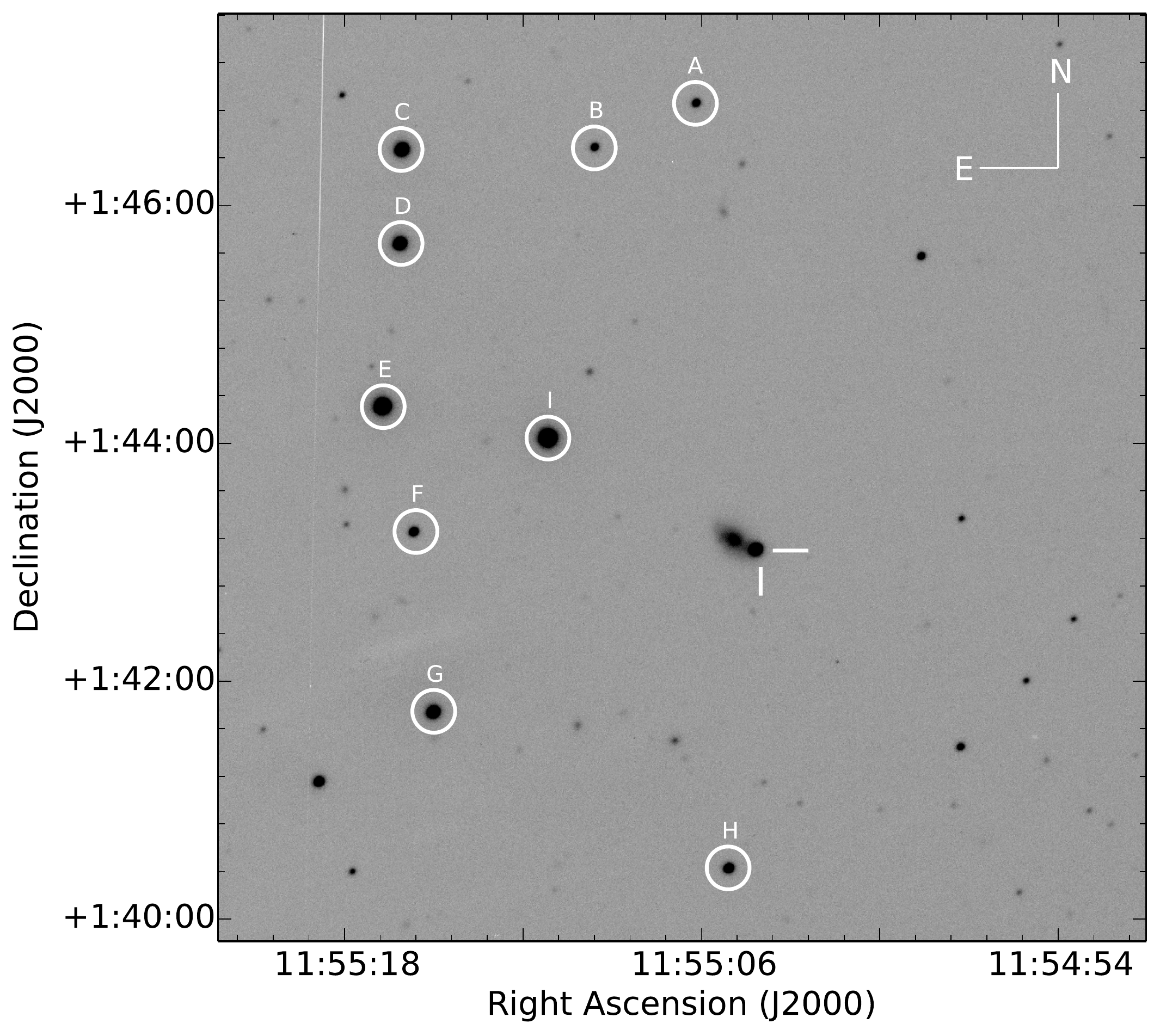}
	\end{center}
	\caption{Local standards in the field of PGC 037392 marked along with SN 2016B. The {\it V}-band image was acquired on 2016 January 07 with the 2.01~m HCT.}
	\label{fig:local_std}
\end{figure}
\begin{table*}
\caption{Summary of the instruments used for the observational campaign of SN 2016B.}
\centering
\smallskip
\flushleft

\begin{tabular}{l l l l l l l}
\hline

Telescope & Location & Instrument  & Pixel Scale    & Imaging & Dispersers/ Grisms   &   \\
                        &  &   & ($''$/pixel) & bands\\
\hline
1.04m Sampurnanand & ARIES Observatory, India & Tek 1k$\times$1k   & 0.53  & {\it VRI}  & - & 1\\
Telescope (ST)         &  &  & & \\
1.3m Devasthal Fast & ARIES Observatory, India  & Andor 512$\times$512 & 0.64  & {\it UBVRI} & - & 2\\
Optical Telescope (DFOT)         &  &  & & \\
1.82m Copernico Telescope & Asiago, Mount Ekar (Italy) & AFOSC & 0.48 & {\it uBVgriz} & Gr04, VPH6 & 3 \\
2.01m Himalayan  & Indian Astronomical   & HFOSC  & 0.296    & {\it UBVRI} & Gr7, Gr8  & 4\\
Chandra Telescope (HCT)         & Observatory, Hanle, India  &  & & \\
2.56m Nordic Optical & Roque de los Muchachos,  & ALFOSC  & 0.19 &  {\it uBVgriz}   & - & 5\\
 Telescope (NOT)        & La Palma, Canaris (Spain) &  & & \\
3.58m Telescopio Nazionale & Roque de los Muchachos, & DOLORES & 0.252 & {\it BVgriz} & LR-B, LR-R & 6\\
 Galileo (TNG)        & La Palma, Canaris (Spain) &  & & \\
 3.6m Devasthal & ARIES Observatory, India & 4k$\times$4k   & 0.19  & {\it R}  & - & 7\\
Optical Telescope (DOT)         &  &  & (2$\times$2 bin) & \\
\hline                                   
\end{tabular}

\label{tab:details_instrument_detectors}      
\end{table*}

\begin{table*}
\caption{{\it Swift} UVOT photometry.}
\centering
\smallskip
\begin{tabular}{c c c c c c c c c}
\hline
UT Date         & JD & Phase$^\dagger$          &   {\textit{uvw2}}    &  {\textit{uvm2}}  & {\textit{uvw1}} & {\textit{uvu}} &  {\textit{uvb}} & {\textit{uvv}} \\
   (yyyy-mm-dd)  &     2457000+        & (days)           & (mag)                 &   (mag)            & (mag)  & (mag)                 &   (mag)            & (mag)\\
\hline

2016-01-03.5 & 391.1 & 9.1   & 13.97 $\pm$ 0.05 & 13.87 $\pm$ 0.06 & 13.49 $\pm$ 0.04 & 13.57 $\pm$ 0.04 & 14.82 $\pm$ 0.05 & 14.81 $\pm$ 0.06 \\
2016-01-06.3 & 393.8 & 11.8 & 14.53  $\pm$ 0.06 & 14.15 $\pm$ 0.05 & 13.87 $\pm$ 0.05 & 13.70 $\pm$ 0.04 & 14.81 $\pm$ 0.05 & 14.88 $\pm$ 0.06 \\
2016-01-07.8 & 395.3 & 13.3 & 14.71 $\pm$ 0.06  & 14.56 $\pm$ 0.06 & 14.06 $\pm$ 0.05 & 13.76 $\pm$ 0.04 & 14.82 $\pm$ 0.05 & 14.77 $\pm$ 0.06 \\
2016-01-10.1 & 397.6 & 15.6 & 15.31 $\pm$ 0.07  & 15.10 $\pm$ 0.06 & 14.60 $\pm$ 0.06 & 13.96 $\pm$ 0.04 & 14.85 $\pm$ 0.05 & 14.84 $\pm$ 0.06 \\
2016-01-11.8 & 399.3 & 17.3 & 15.72  $\pm$ 0.08 & 15.53 $\pm$ 0.06 & 14.97 $\pm$ 0.06 & 14.09 $\pm$ 0.05 & 14.89 $\pm$ 0.05 & 14.92 $\pm$ 0.06 \\
2016-01-14.2 & 401.7 & 19.7 & 16.36 $\pm$ 0.09  & 16.15 $\pm$ 0.07 & 15.52 $\pm$ 0.07 & 14.42 $\pm$ 0.05 & 14.95 $\pm$ 0.05 & 14.94 $\pm$ 0.06 \\
2016-01-16.1 & 403.9 & 21.9 & 16.98  $\pm$ 0.11 & 16.76 $\pm$ 0.08 & 16.10 $\pm$ 0.08 & 14.75 $\pm$ 0.05 & 15.09 $\pm$ 0.05 & 14.97 $\pm$ 0.06\\
2016-01-17.6 & 405.1 & 23.1 & 17.29 $\pm$ 0.13  & -  & 16.28 $\pm$ 0.09 & 14.93 $\pm$ 0.06 & 15.18 $\pm$ 0.05 & - \\
\hline                                   
\end{tabular}
\newline
\begin{tablenotes}
\item[a]$^\dagger$since explosion epoch t$_0$ = 2457382.0 JD (2015 December 25.4)
     \end{tablenotes}

\label{uv_photometry}      
\end{table*}
\begin{table*}
\caption{Coordinates and photometry of the local sequence reference stars in the $UBVRI$ bands.}
\begin{tabular}{llcccccc}
\hline
 ID & $ \alpha_{J2000.0} $ & $ \delta_{J2000.0} $ & $U$ & $B$  & $V$  & $R$ &$I$ \\
 & (hh:mm:ss) &(dd:mm:ss) & mag & mag & mag & mag & mag \\
\hline 
A & 11:55:06.2 & +01:46:51.5 & 17.380 (.045) & 17.371 (.009) & 16.760 (.009) & 16.375 (.010) & 16.037 (.009) \\
B & 11:55:09.6 & +01:46:29.0 & 19.181 (.096) & 17.974 (.012) & 16.849 (.009) & 16.235 (.010) & 15.695 (.006) \\
C & 11:55:16.1 & +01:46:28.2 & 15.212 (.038) & 14.826 (.007) & 14.202 (.008) & 13.837 (.010) & 13.524 (.007) \\
D & 11:55:16.1 & +01:45:40.8 & 14.984 (.038) & 14.892 (.007) & 14.410 (.007) & 14.085 (.010) & 13.808 (.006) \\
E & 11:55:16.7 & +01:44:18.5 & 14.451 (.039) & 14.027 (.007) & 13.380 (.007) & 13.024 (.009) & 12.725 (.006) \\
F & 11:55:15.6 & +01:43:15.5 & 17.026 (.042) & 16.612 (.025) & 16.055 (.008) & 15.676 (.009) & 15.362 (.008) \\
G & 11:55:15.0 & +01:41:44.8  & 15.507 (.040) & 15.156 (.051) & 14.621 (.007) & 14.268 (.009) & 13.961 (.007) \\
H & 11:55:05.1 & +01:40:25.8  & 17.330 (.047) & 16.717 (.007) & 15.822 (.008) & 15.341 (.009) & 14.900 (.005) \\
\hline

\end{tabular}
\label{tab:local}   
\end{table*}

\begin{table*}
 \caption{Optical photometry of SN 2016B.}
\centering
\smallskip
\begin{tabular}{c c c c c c c c c}
\hline
UT Date         & JD & Phase$^\dagger$          &   {\textit{U}}    &  {\textit{B}}  & {\textit{V}} & {\textit{R}} &  {\textit{I}} & {\textit{Tel}} \\
   (yyyy-mm-dd)  &     2457000+        & (days)           & (mag)                 &   (mag)            & (mag)  & (mag)                 &   (mag)            & \\
\hline
2016-01-07.9 & 395.4 & 13.4   & -                                    & 14.65 $\pm$ 0.02 & 14.73 $\pm$ 0.03 & 14.56 $\pm$ 0.03 & 14.48 $\pm$ 0.01 & HCT \\
2016-01-12.9 & 400.4 & 18.4   & 14.40  $\pm$ 0.07  & 14.76 $\pm$ 0.02 & 14.78 $\pm$ 0.03 & 14.56 $\pm$ 0.04 & 14.47 $\pm$ 0.01 & HCT \\
2016-01-12.9 & 400.5 & 18.5   & 14.44  $\pm$ 0.07  & 14.76 $\pm$ 0.02 & 14.70 $\pm$ 0.03 & 14.51 $\pm$ 0.04 & 14.51 $\pm$ 0.01 & DFOT \\
2016-01-13.9 & 401.4 & 19.4   & 14.54  $\pm$ 0.07  & 14.77 $\pm$ 0.02 & 14.66 $\pm$ 0.03 & 14.51 $\pm$ 0.04 & 14.50 $\pm$ 0.01 & DFOT \\
2016-01-15.9 & 403.5 & 21.5   & 14.76  $\pm$ 0.05  & 15.13 $\pm$ 0.02 & 14.74 $\pm$ 0.03 & 14.55 $\pm$ 0.04 & 14.52 $\pm$ 0.01 & DFOT \\
2016-01-18.9 & 406.4 & 24.4   &  -                                   & -                                  & 14.74 $\pm$ 0.03 & 14.55 $\pm$ 0.03 & 14.51 $\pm$ 0.01 & DFOT \\
2016-01-19.2 & 406.6 & 24.6   & 14.86  $\pm$ 0.04  & 14.97 $\pm$ 0.02 & 14.77 $\pm$ 0.03 & 14.59 $\pm$ 0.02 & 14.54 $\pm$ 0.02 & EKAR \\
2016-01-23.8 & 411.3 & 29.3   & 15.50  $\pm$ 0.07  & 15.38 $\pm$ 0.02 & 14.94 $\pm$ 0.03 & 14.65 $\pm$ 0.04 & 14.52 $\pm$ 0.01 & HCT \\
2016-01-30.8 & 418.3 & 36.3   & -                                    & -                                  &       -                            & 14.73 $\pm$ 0.03 & 14.59 $\pm$ 0.01 & ST \\
2016-01-31.8 & 419.3 & 37.3   & -                                    & 15.77 $\pm$ 0.02 & -                                  & -                                  & -                                  & DFOT \\
2016-01-31.9 & 419.4 & 37.4   & -                                    & -                                  & 15.00 $\pm$ 0.03 & 14.73 $\pm$ 0.03 & 14.59 $\pm$ 0.01 & ST \\
2016-02-02.8 & 421.3 & 39.3   & -                                    & -                                  &       -                            & 14.73 $\pm$ 0.02 & 14.64 $\pm$ 0.01 & ST \\
2016-02-08.9 & 427.4 & 45.4   & -                                    & -                                  &       -                            & 14.75 $\pm$ 0.04 & 14.68 $\pm$ 0.01 & DFOT \\
2016-02-10.8 & 429.4 & 47.4   & 16.89 $\pm$ 0.04   & 16.02 $\pm$ 0.02 & 15.20 $\pm$ 0.03 &  -                                 & -                                  & DFOT \\
2016-02-10.8 & 429.4 & 47.4   & -                                    & -                                  & -                                  & 14.83 $\pm$ 0.03 & 14.66 $\pm$ 0.01 & ST \\
2016-02-11.8 & 430.4 & 48.4   & 16.88 $\pm$ 0.05   & 16.04 $\pm$ 0.02 & 15.23 $\pm$ 0.03 & 14.86 $\pm$ 0.02 & 14.70 $\pm$ 0.01 & DFOT \\
2016-02-12.9 & 431.4 & 49.4   & 16.83 $\pm$ 0.06   & 15.98 $\pm$ 0.02 & 15.21 $\pm$ 0.03 & 14.84 $\pm$ 0.03 & 14.69 $\pm$ 0.01 & DFOT \\
2016-02-13.9 & 432.4 & 50.4   & 17.04 $\pm$ 0.04   & 16.10 $\pm$ 0.02 & 15.28 $\pm$ 0.03 & 14.90 $\pm$ 0.03 & 14.73 $\pm$ 0.01 & DFOT \\
2016-02-14.9 & 433.5 & 51.5   & -                                    & -                                  & 15.44 $\pm$ 0.03 & -                                 & 14.71 $\pm$ 0.04  & ST \\
2016-02-20.9 & 439.4 & 57.4   & 17.14 $\pm$ 0.07   & 16.38 $\pm$ 0.02 & 15.43 $\pm$ 0.03 & 14.99 $\pm$ 0.04 & 14.76 $\pm$ 0.01 & HCT \\
2016-02-28.8 & 447.4 & 65.4   & -                                    & 16.52 $\pm$ 0.02 & 15.50 $\pm$ 0.03 & 15.04 $\pm$ 0.04 & 14.82 $\pm$ 0.01 & HCT \\
2016-02-29.8 & 448.4 & 66.4   & -                                    & -                                  & 15.58 $\pm$ 0.04 & 15.00 $\pm$ 0.04 & 14.81 $\pm$ 0.01 & ST \\
2016-03-01.8 & 449.4 & 67.4   & -                                    & -                                  &       -                            & 15.02 $\pm$ 0.04 & 14.82 $\pm$ 0.01 & ST \\
2016-03-02.0 & 449.5 & 67.5   & -                                    & 16.68 $\pm$ 0.02 & 15.52 $\pm$ 0.03 &  & & NOT \\
2016-03-02.8 & 450.3 & 68.3   & -                                    & -                                  & 15.65 $\pm$ 0.03 & 15.01 $\pm$ 0.04 &14.82 $\pm$ 0.01 & ST \\
2016-03-07.8 & 455.3 & 73.3   & -                                    & 16.59 $\pm$ 0.02 & -                                  &  -                                 & & DFOT \\
2016-03-10.9 & 458.4 & 76.4   & 17.84 $\pm$ 0.09   & 16.69 $\pm$ 0.03 & 15.49 $\pm$ 0.04 & 15.10 $\pm$ 0.01 & & EKAR \\
2016-03-15.8 & 463.3 & 81.3   & -                                    & -                                  & 15.73 $\pm$ 0.04 & 15.01 $\pm$ 0.04 & 14.83 $\pm$ 0.01 & ST \\
2016-03-18.9 & 466.4 & 84.4   & -                                    & -                                  & 15.82 $\pm$ 0.04 & 15.09 $\pm$ 0.04 & 14.90 $\pm$ 0.01 & ST \\
2016-03-31.8 & 479.3 & 97.3   & 17.88 $\pm$ 0.07   & 16.81 $\pm$ 0.02 & 15.65 $\pm$ 0.03 & 15.12 $\pm$ 0.04 & 14.90 $\pm$ 0.01 & HCT \\
2016-04-07.9 & 486.4 & 104.4  & -                                  &  -                                 & 15.72 $\pm$ 0.03 & 15.19 $\pm$ 0.03 & 14.96 $\pm$ 0.01 & HCT \\
2016-04-28.7 & 507.2 & 125.2  & -                                  & 17.54 $\pm$ 0.02 & 16.12 $\pm$ 0.03 & 15.51 $\pm$ 0.04 & 15.23 $\pm$ 0.01 & HCT \\
2016-05-02.7 & 511.2 & 129.2  & -                                  & -                                  & 16.33 $\pm$ 0.04 & 15.64 $\pm$ 0.03 & 15.42 $\pm$ 0.01 & ST \\
2016-05-25.9 & 534.5 & 152.5  & -                                  & 19.31 $\pm$ 0.03 & 18.10 $\pm$ 0.05 &       -                            & & TNG \\
2016-05-26.7 & 535.2 & 153.2  & -                                 & -                                   &     -                              & 16.87 $\pm$ 0.03 & 16.45 $\pm$ 0.02 & ST \\
2016-06-30.6 & 570.1 & 188.1  & -                                 & -                                   & 18.10 $\pm$ 0.04 & 17.26 $\pm$ 0.04 & 16.74 $\pm$ 0.02 & HCT \\
2017-04-02.9 & 847.0 & 465.0  & -                                 & -                                   & -                                  & 19.79 $\pm$ 0.05 & - & DOT \\ 
\hline                                   
\end{tabular}
\newline
\begin{tablenotes}
\item[a]$^\dagger$since explosion epoch t$_0$ = 2457382.0 JD (2015 December 25.4)
     \end{tablenotes}
\label{photometry}
\end{table*}

The UV photometry was obtained from {\it Swift} Optical/Ultraviolet Supernova Archive \citep[SOUSA; https://archive.stsci.edu/prepds/sousa/;][]{2014Ap&SS.354...89B}. The reduction is based on that of \cite{2009AJ....137.4517B}, including subtraction of the host galaxy count rates and uses the revised UV zeropoints and time-dependent sensitivity from \cite{2011AIPC.1358..373B}. The {\it Swift} UVOT magnitudes are given in Table~\ref{uv_photometry}.

We performed point spread function (PSF) fitting photometry on the optical data using DAOPHOT II \citep{1987PASP...99..191S}, pre-processing the images with bias subtraction, flat fielding and cosmic ray removal \citep[{\sc L.A.Cosmic},][]{2001PASP..113.1420V}. Since SN~2016B exploded in the outskirts of the host galaxy, a careful selection of the aperture for PSF photometry is sufficient to eliminate any host galaxy contribution to the SN flux. The first aperture used for photometry is the mean FWHM of the frame, while the sky flux is estimated in an 8~pixel wide annular region with an inner radius which is four times the FWHM.

Further, to convert the instrumental magnitudes to the standard magnitudes, a set of local standard stars were generated in the SN field. To this end, observations of three Landolt standard fields \citep{2009AJ....137.4186L}, namely, PG 0918, PG 1047 and PG 1323, were carried out on 2016 Feb 28 under photometric night conditions (median FWHM seeing in $V$ band $\sim$ 2$^{\prime\prime}$) with the 2.01m Himalayan Chandra Telescope~(HCT) at Indian Astronomical Observatory~(IAO), Hanle. These fields were observed at three altitudes covering airmass from 1.176 to 1.773. The $V$-band magnitudes of the stars in the standard field lie in the range 12.08 to 14.49~mag and the $B-V$ colour varies between $-$0.290 and 1.044~mag. With the aid of the instrumental and standard magnitudes of the Landolt field stars, the zero points, colour coefficients and extinction values of the transformation equations were fitted using the least square linear regression technique as described in \cite{1992JRASC..86...71S} and the values thus obtained are given below:
$$ u = U + (3.40 \pm 0.27) + (0.08 \pm 0.13)(B-V) + (0.36 \pm 0.16)X$$
$$ b = B + (1.28\pm 0.02) + (-0.004 \pm 0.020)(B-V) + (0.20 \pm 0.02)X$$
$$ v = V + (0.71 \pm 0.03) + (-0.08 \pm 0.04)(B-V) + (0.12 \pm 0.03)X$$
$$ r = R + (0.79 \pm 0.05) + (-0.16 \pm 0.08)(V-R) + (0.08 \pm 0.05)X$$
$$ i  = I + (1.026 \pm 0.005) + (-0.027 \pm 0.004)(V-I) + (0.055 \pm 0.005)X$$
where X is the airmass. The rms scatter between the observed and standard magnitudes of the Landolt field stars are 0.08 mag in $U$, 0.03 mag in $V$ and 0.02 mag in $BRI$ bands. We used these coefficients to generate a photometric sequence of eight non-variable stars in the SN field. The eight local standards are marked in Fig. \ref{fig:local_std} and their magnitudes are tabulated in Table~\ref{tab:local}. The final SN magnitudes (listed in Table \ref{photometry}) are then obtained differentially with respect to the local standards by estimating the nightly zero points. The errors in the magnitudes are the combined PSF fitting and photometric calibration errors propagated in quadrature.

\section{Spectroscopy}
\label{spectro}
We removed the detector signatures from the frames using standard tasks in {\sc IRAF}. The one-dimensional spectra extracted using the $APALL$ task, were wavelength and flux calibrated using arc lamps and spectrophotometric standard star spectra, respectively, obtained at a similar airmass on the same night or a close-by night. To validate the wavelength calibration, night sky emission lines were used, and shifts were applied when necessary. A scaling factor to match the spectroscopic and photometric continuum flux, is multiplied to the calibrated spectra to account for slit losses. Finally each spectrum was de-redshifted to the host galaxy rest frame using the $DOPCOR$ task.
\begin{table*}
\caption{Log of the spectroscopic observations.}
\centering
\smallskip
\begin{tabular}{c c c c c c c c}
\hline \hline
Date         & Phase$^\dagger$          & Grism      & Spectral Range        & Resolution & Instrument & Telescope       \\
                  & (Days)           & & (\AA)                                &   & \\
\hline
2016-01-07      & 13.4 & Gr7, Gr8 & 3800-6840, 5800-9000 & 1330, 2190 & HFOSC & HCT \\
2016-01-11      &  17.5 & Gr7, Gr8 & 3800-6840, 5800-9000 & 1330, 2190 & HFOSC & HCT \\
2016-01-13      &  19.4 & Gr7, Gr8 & 3800-6840, 5800-9000 & 1330, 2190 & HFOSC & HCT \\
2016-01-19      &  24.6 &  Gr04      & 3360-7740       & 311 & AFOSC & EKAR \\
2016-01-20      &  26.4 & Gr7, Gr8 & 3800-6840, 5800-9000 & 1330, 2190 & HFOSC & HCT \\
2016-01-22      &  28.4 & Gr7, Gr8 & 3800-6840, 5800-9000 & 1330, 2190 & HFOSC & HCT \\
2016-01-30      &  36.3 & Gr7, Gr8 & 3800-6840, 5800-9000 & 1330, 2190 & HFOSC & HCT \\
2016-02-05      &  42.2 & Gr7, Gr8 & 3800-6840, 5800-9000 & 1330, 2190 & HFOSC & HCT \\
2016-02-20      &  57.3 & Gr7, Gr8 & 3800-6840, 5800-9000 & 1330, 2190 & HFOSC & HCT \\
2016-02-21      &  58.2 & Gr7, Gr8 & 3800-6840, 5800-9000 & 1330, 2190 & HFOSC & HCT  \\
2016-02-28      &  65.3 & Gr7, Gr8 & 3800-6840, 5800-9000 & 1330, 2190 & HFOSC & HCT \\
2016-03-04      &  70.3 & Gr7, Gr8 & 3800-6840, 5800-9000 & 1330, 2190 & HFOSC & HCT \\
2016-03-10      &  76.4 &  Gr04      & 3360-7740       & 311 & AFOSC & EKAR  \\
2016-03-31      &  97.3 & Gr7, Gr8 & 3800-6840, 5800-9000 & 1330, 2190 & HFOSC & HCT \\
2016-04-27      & 124.4 & VPH6 & 4500-10000 & 500 & AFOSC & EKAR \\
2016-04-28      & 125.2 & Gr7, Gr8 & 3800-6840, 5800-9000 & 1330, 2190 & HFOSC & HCT \\
2016-05-25      & 152.5 & LR-B, LR-R & 3000-8430, 4470-10073 & 585, 714 & DOLORES & TNG\\
2016-06-19      & 177.1 & Gr7, Gr8 & 3800-6840, 5800-9000 & 1330, 2190 & HFOSC & HCT \\
\hline                                   
\end{tabular}
\newline
\begin{tablenotes}
\item[a]$^\dagger$since explosion epoch t$_0$ = 2457382.0 JD (2015 December 25.4)
     \end{tablenotes}

\label{tab:spectra_log}      
\end{table*}

\bsp	
\label{lastpage}
\end{document}